\documentclass{aa} 
\usepackage{graphicx} 
\usepackage{amsmath} 
\usepackage{natbib}

\bibpunct{(}{)}{;}{a}{}{,}
\usepackage{txfonts} 
\begin{document} 
\title{Constraining dark energy via baryon acoustic oscillations in the
  (an)isotropic light-cone power spectrum}  

   \author{Christian Wagner 
          \and 
          Volker M\"uller
\and
Matthias Steinmetz
          } 
 
   \offprints{C. Wagner, \email{cwagner@aip.de}} 
 
\institute{AIP - Astrophysikalisches Institut Potsdam, An der Sternwarte 16, 
D-14482 Potsdam, Germany }

   \date{Received / Accepted} 

\authorrunning{Wagner et al.}
\titlerunning{Constraining DE via BAO in the (an)isotropic light-cone power 
spectrum} 
  
\abstract 
{The measurement of the scale of the baryon acoustic oscillations (BAO) in 
the galaxy power spectrum as a function of redshift is a promising method to 
constrain the equation-of-state parameter of the dark energy $w$.} 
{To measure the scale of the BAO precisely, a substantial volume of space must be surveyed. We test whether light-cone effects are important and
whether the scaling relations used to compensate for an incorrect reference
cosmology are in this case sufficiently accurate. We investigate the 
degeneracies in the cosmological parameters and the benefits of 
using the two-dimensional anisotropic power spectrum. Finally, we estimate 
the uncertainty with which $w$ can be measured by 
proposed surveys at redshifts of about $z=3$ and $z=1$, respectively.} 
{Our data is generated by cosmological N-body simulations of the standard 
$\Lambda$CDM scenario. We construct galaxy catalogs by ``observing'' the 
redshifts of different numbers of mock galaxies on a light cone at redshifts 
of about $z=3$ and $z=1$. From the ``observed'' redshifts, we calculate the
distances, assuming a reference cosmology that depends on $w_{\rm ref}$. We do this for 
$w_{\rm ref}=-0.8, -1.0$, and $-1.2$  holding the other cosmological parameters 
fixed. By fitting the corresponding (an)isotropic power spectra, we determine 
the apparent scale of the BAO and the corresponding $w$.}
{In the simulated survey we find that light-cone effects are small and
that the simple scaling relations used to correct for the cosmological
distortion work fairly well even for large survey volumes. 
The analysis of the two-dimensional anisotropic power spectra enables an 
independent determination to be made of the apparent scale of the BAO, perpendicular and
parallel to the line of sight. This is essential for two-parameter $w$-models, such 
as the redshift-dependent dark energy model $w=w_0+(1-a) w_a$.
Using Planck priors for the matter and baryon density and $\Delta H_0=5\%$ for the Hubble constant,
we estimate that the BAO measurements of future surveys 
around $z=3$ and $z=1$ will be able to constrain, independently of other
cosmological probes, a constant $w$ to $\sim 12\%$ and $\sim 11\%$ (68\% c.l.), respectively.
} 
{}
 
 
\maketitle

\section{Introduction} 
There are two main observational indications of the existence of an unknown
energy component in the Universe, which is usually referred to as dark energy (DE). First, 
observations of distant supernovae Ia \citep{riess, perlmutter} favor an accelerating
expansion of the Universe and therefore imply an energy component of
negative pressure $P$. In particular, the parameter $w$ of the equation of state 
(EOS) $P=w\rho$ must obey $w<-1/3$. Second, measurements of the anisotropies 
in the cosmic microwave background (CMB) \citep{wmap1,wmap3} in combination with observations of 
the large-scale structure of the Universe argue for a 
spatially flat Universe. Matter (baryonic and dark) contributes however less 
than 30\% to the critical density. Hence, about 70\% of the present-day 
energy density of the Universe appears to be in an unknown form of energy. 
The simplest way to account for this missing energy and the accelerating 
expansion is to introduce a cosmological constant $\Lambda$ in Einstein's 
equations, which has a redshift-independent EOS parameter  $w=-1$. So far 
all observations appear to agree with the model of a cosmological constant. 
There is however at least one theoretical drawback. The observed value of
$\Lambda$, which is interpreted as vacuum energy, is highly inconsistent with
current predictions by particle physics, a discrepancy commonly referred to as
the cosmological constant 
problem (for a review, see \citet{cc}). This has motivated consideration 
of more general dark energy models that have a redshift-dependent EOS 
parameter $w(z)$. The measurement of the parameter $w$ can therefore
help to distinguish between not only a simple cosmological constant and other dark 
energy models, but potentially also between these different models.

Several possible methods to constrain the EOS parameter $w$ are summarized by
the Dark Energy Task Force Report \citep{DETF_report}. 
The method that we consider here uses the expansion history 
of the Universe. To measure this precisely, we require a
standard candle or a standard ruler. For redshifts up to $z \sim 1.8$, 
supernovae Ia can be observed and calibrated to be standard candles, with which one can measure
the luminous distance $D_L(z)$. \citet{cosmiccomp_1,cosmiccomp_2} 
proposed that the baryon acoustic oscillations (BAO) imprinted in the galaxy power 
spectrum could be utilized as a standard ruler. The BAO have the same origin as the acoustic
peaks in the anisotropies of the CMB. Before recombination, baryons and photons 
were tightly coupled. Gravitation and radiation pressure 
produced acoustic oscillations in this hot plasma; during the expansion of the
Universe, the plasma then cooled and finally nuclei and electrons recombined. The released
photons propagated through the expanding Universe and are observed by
ourselves as 
the highly redshifted radiation of the CMB; in contrast, the baryons followed
the clustering of the dark matter and eventually collapsed to form
galaxies. Since the dark matter did not participate 
in the acoustic oscillations, the oscillatory feature in the galaxy power 
spectrum is far less pronounced than for the CMB photons 
\citep{peebles, sunyaev, bond, holtzman, sugiyama, ehu98}. 

When the physical scale of the BAO has been calibrated using precise CMB
measurements, it can be applied as a standard ruler for measuring the angular 
diameter distance $D_A(z)$ and Hubble parameter $H(z)$.
In contrast to supernovae Ia, the scale of the BAO is a more reliable 
standard ruler at high redshifts. As the number of unperturbed peaks
and troughs corresponding to the BAO in the power spectra increases, 
the wavelength of the BAO can be determined more accurately. 
On scales where structure growth is already nonlinear, the oscillations 
cannot be easily discerned \citep{millenium, marenostrum, unwiggle}. Hence, with
decreasing redshift the uncertainty in the observed scale of the BAO
increases. To have good statistics for the first peaks a large volume has to
be surveyed. The BAO have been detected in the present-day largest galaxy
redshift surveys \citep{peak,cole}. Data sets studying larger volumes and higher redshifts are
however required to achieve tight
constraints on dynamical DE models. Galaxy redshift surveys at about
$z=3$, $1$, or $0.5$, like HETDEX \citep{hetdex}, the WFMOS BAO survey \citep{glazebrook_wp,wfmos}, and BOSS \citep{boss}, are designed for measuring
the scale of the BAO and the EOS parameter $w$ with a precision to 
a few percent and thereby constrain a variety of DE models.

Many studies presented methods for extracting the scale of the BAO and 
estimated the accuracy of its measurement achievable by future surveys. In
these analyses, Monte Carlo simulations 
\citep{blake_2003,blake_2,universal_fitting}, Fisher matrix techniques 
\citep{linder,rings,seo_2003,matsubara,huetsi_sz,seo_2007}, N-body simulations 
\citep{angulo,white,seo_2005,huff,koehler,angulo2007}, and observational data 
\citep{peak,huetsi_astro_ph,huetsi_sdss,huetsi_constraints} were used. 

In this article, we attempt to include all important (physical and observational) 
effects for measuring the BAO and deriving constraints 
on the EOS parameter $w$. Our perspective is that of an observer, i.e.~the 
starting point for the BAO measurement should be a galaxy catalog that 
provides celestial coordinates and redshifts of the galaxies. Since we do not have in hand
the type of real observations that we require, we have first to generate mock
catalogs; we achieve this by completing ``observations'' on the data products of N-body simulations. Using these ``observations'', we are able to study for 
the first time in detail the light-cone 
effect and the accuracy of the scaling relations used to compensate the 
cosmological distortion that arises by assuming an incorrect reference cosmology. 
A crucial point for all BAO measurements is the fitting method. We develop 
a method which uses only the oscillatory part of the power spectrum 
in a way that produces unbiased and robust results.
Further, we compare the results of fitting the
angle-averaged one-dimensional power spectrum and anisotropic 
two-dimensional power spectrum. Finally, we predict the 
uncertainty with which proposed surveys will be able to measure the EOS 
parameter $w$ by assuming two different $w$-models.

The paper is structured as follows. In Sect.~2, we review how the EOS
parameter $w$ is measured using BAO. In Sect.~3, we explain how we generate 
the ``observed'' data from N-body simulations, and in Sect.~4, we describe the 
power spectrum calculation and our fitting method. In Sect.~5, we present our 
results, and finally we provide our conclusions in Sect.~6.

\section{Cosmology with baryon acoustic oscillations}
\label{cosmography} 
Knowledge of the true comoving scales, both parallel ($r_\parallel$) and transverse 
($r_\bot$) to the line of sight, of an observed physical property at a given
redshift $z$, in the case of BAO statistical properties of large-scale
structure, 
enables us to derive the Hubble parameter $H(z)$ and angular diameter 
distance $D_A(z)$ from the measured quantities (redshift $\Delta z$ and 
angle $\Delta \theta$) :
\begin{align}
\begin{aligned}
r_\parallel&=\frac{c\Delta z}{H(z)}\,,\\
r_\bot&=(1+z)D_A(z)\Delta \theta\,.
\end{aligned}
\end{align}
In a flat Universe and for moderate redshifts at which the contribution of
radiation can be neglected, the Hubble parameter depends on three cosmological 
parameters, the present-day Hubble parameter $H_0$, fraction of matter $\Omega_{\rm m}$, 
and the dark energy EOS parameter $w$(z), in the following manner
\begin{equation}
\label{hubble}
H(z)=H_0\sqrt{\Omega_{\rm m}(1+z)^3+(1-\Omega_{\rm m}) 
   \exp \left ( 3\int_0^z {\frac{1+w(\tilde z)}{1+\tilde z}\,
      {\rm d} \tilde z}\right)} \,.
\end{equation}
The angular diameter distance $D_A(z)$ is a functional of the Hubble parameter 
and is given by
\begin{equation}
D_A(z)=\frac{c}{1+z}\int_0^z{\frac{{\rm d}\tilde z}{H(\tilde z)}}\,.
\end{equation}
Assuming that we know apart from $w$ all important cosmological parameters 
to significant precision from other 
observations, especially from CMB precision measurements such as WMAP
\citep{wmap3} and direct measurements of the Hubble constant $H_0$ by the
HST Key project \citep{HST}, one can constrain the EOS parameter $w$ using this chain of 
equations. Since $w$ appears in an integral over $z$ in Eq. (\ref{hubble}), 
observations at a single redshift cannot be used to measure the value of $w$ at 
this redshift. A certain model for $w$ has to be assumed before its value can be
measured. In this paper, we assume a constant
$w$ and a simple redshift-dependent parameterization 
$w=w_0 + (1-a) w_a$; this parameterization is used frequently in the
literature (compare Dark Energy Task Force Report \citep{DETF_report}) and advocated by \citet{wa_model} 
to be robust and applicable to a wide range of dark energy models.

One problem of using the scale of the BAO as a standard ruler is that the BAO appear in a 
statistical quantity. We cannot measure $\Delta z$ of the BAO, but we can measure 
the redshifts of the galaxies and then, assuming a reference cosmology, 
reconstruct their positions and derive their power spectrum. From this power 
spectrum, we can determine the apparent scale of the BAO and compare this 
with its true value. If they agree, we have used the correct cosmology. In
principle, for every trial cosmology we have to recalculate the distances and 
recompute the power spectrum. A more efficient method is to scale 
appropriately the power spectrum derived for the reference cosmology using 
the following approximations \citep{seo_2003,blake_2}
\begin{align}
\begin{aligned}
k_\parallel=\frac{H(z)}{H^{\rm ref}(z)} k_\parallel^{\rm ref}\,,\quad \label{approx}
k_\bot=\frac{D_A^{\rm ref}(z)}{D_A(z)}k_\bot^{\rm ref} \,.
\end{aligned}
\end{align}
In this paper, we test if these approximations are still sufficiently accurate for 
very large surveys where $z$ varies substantially within the survey. 
We achieve this by using three reference cosmologies, which differ only in the EOS 
parameter $w_{\rm ref}=-0.8,\,-1.0,\,-1.2$. We use the relation (\ref{approx}) 
to scale the derived power spectra and compare the results. For convenience, we
define the following scaling factors parallel and transverse to the line of 
sight and an isotropic one for the one-dimensional angle-averaged power spectrum
\begin{align}
\begin{aligned}
\lambda_\parallel&=\frac{H(z)}{H^{\rm ref}(z)}\,,\quad \label{scalingfactors}
\lambda_\bot=\frac{D_A^{\rm ref}(z)}{D_A(z)} \,,\\
\lambda_{\rm iso}&=(\lambda_\bot^2 \lambda_\parallel)^{1/3}\,.
\end{aligned}
\end{align}

\begin{figure}[htbp]
\centering
\includegraphics[angle=-90,width=4.35cm]{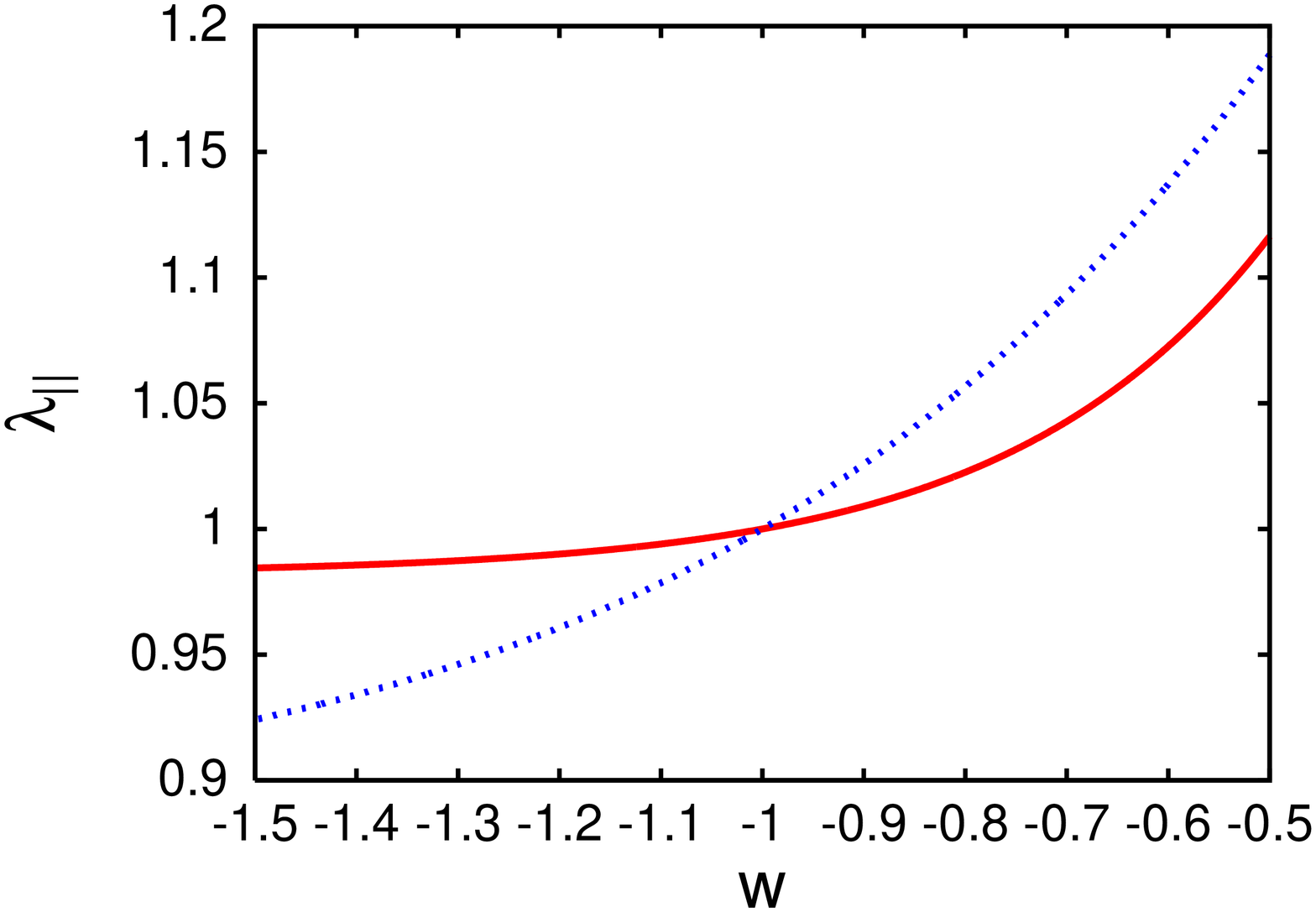}
\includegraphics[angle=-90,width=4.35cm]{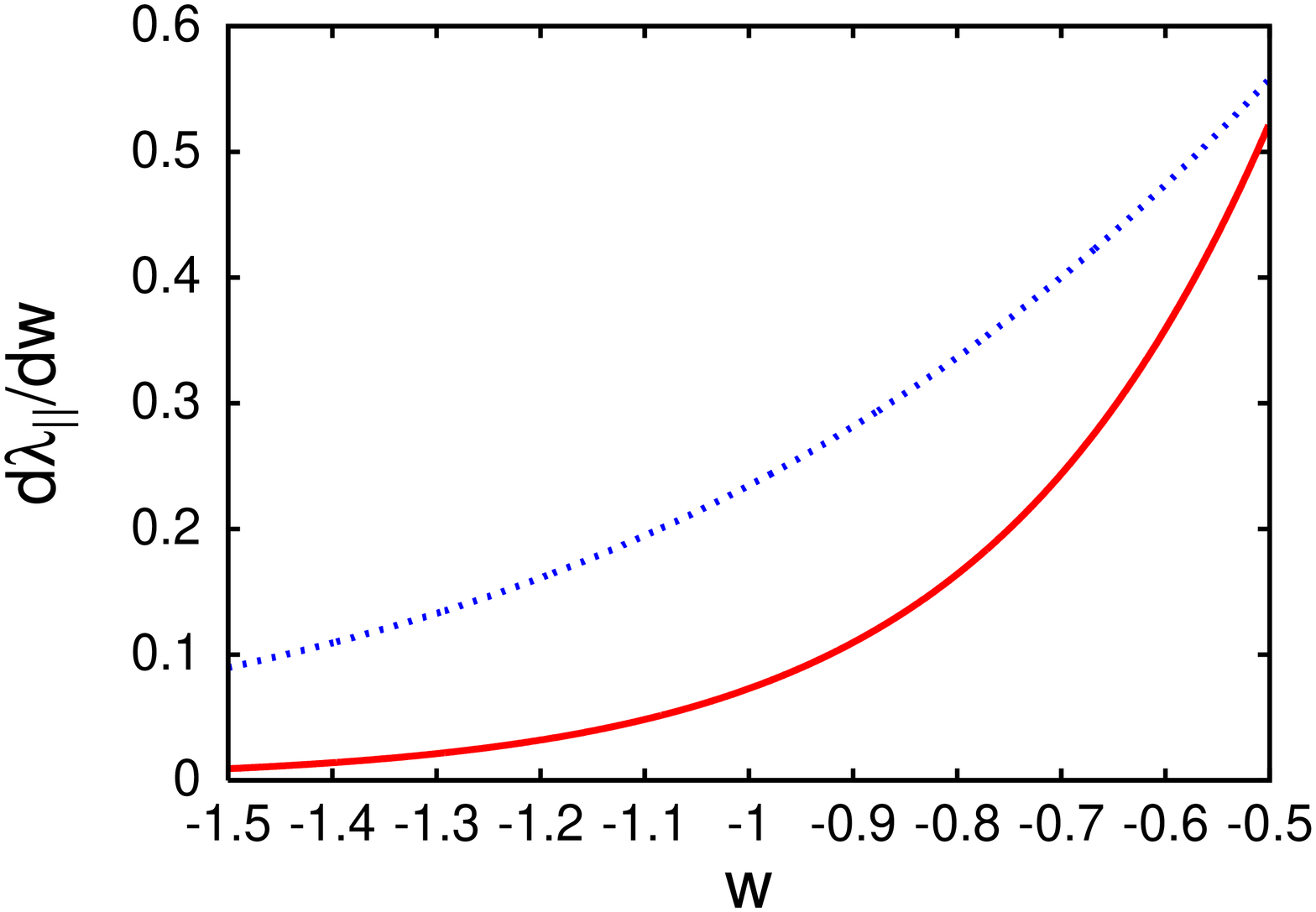}
\includegraphics[angle=-90,width=4.35cm]{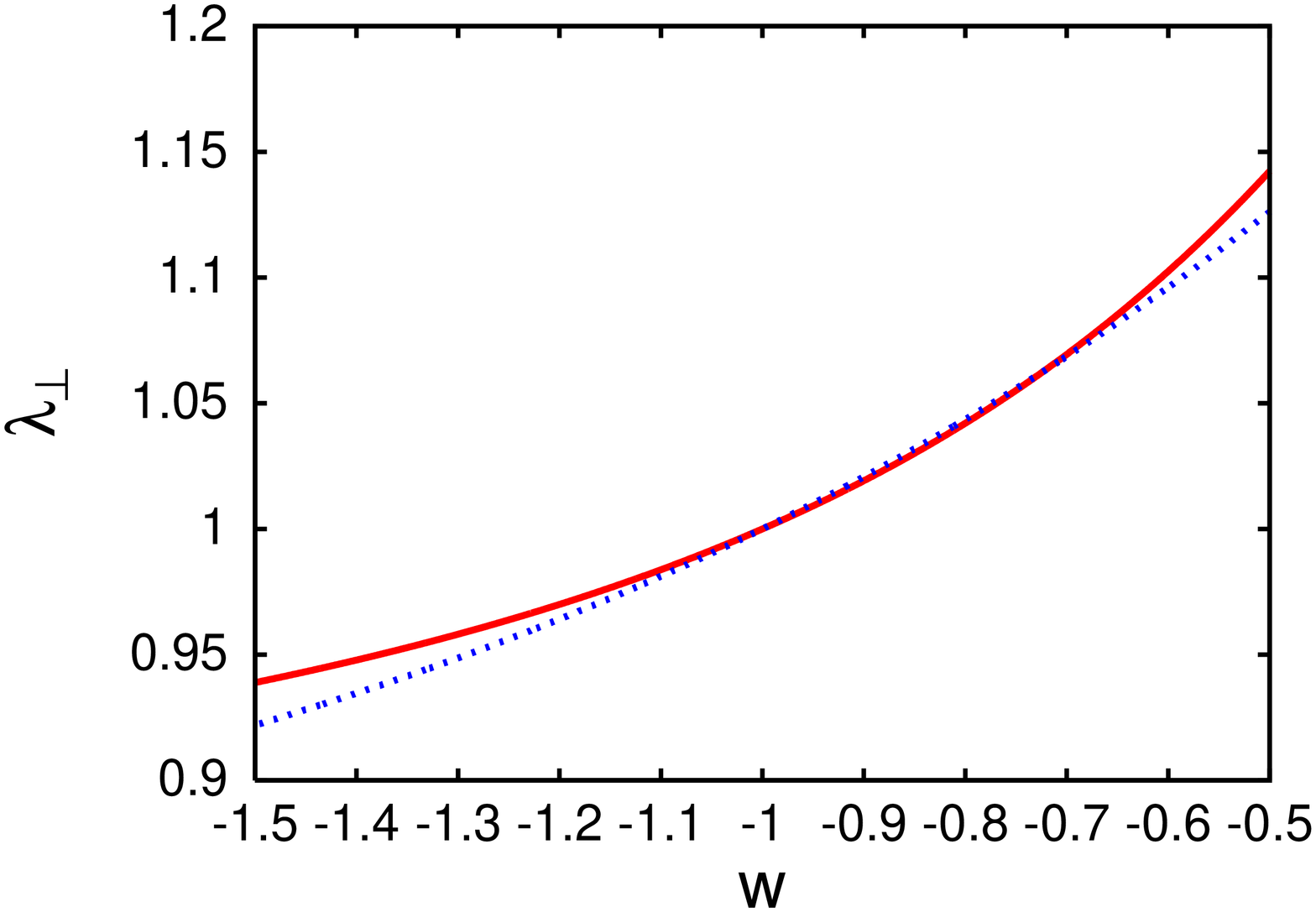}
\includegraphics[angle=-90,width=4.35cm]{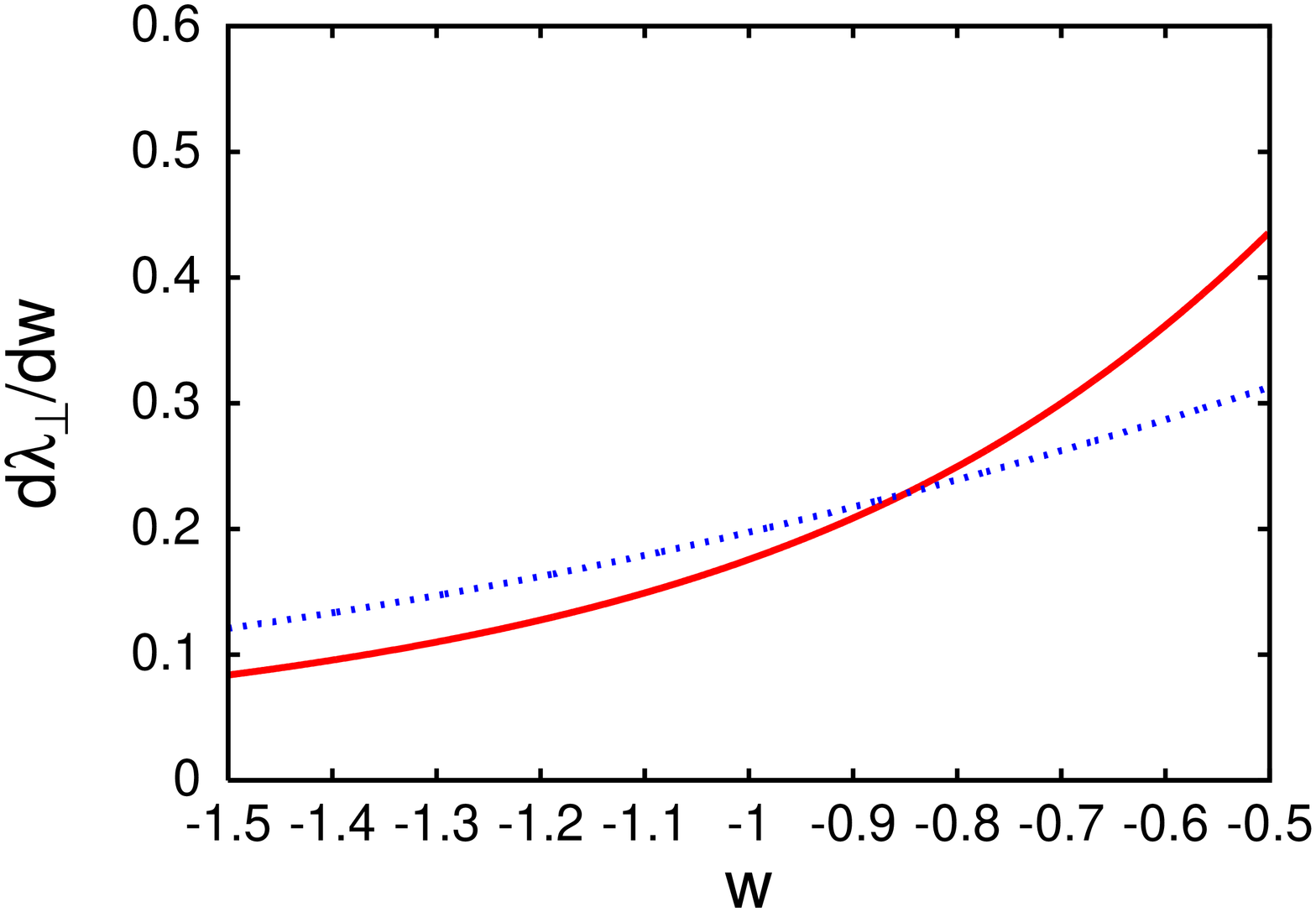}
\includegraphics[angle=-90,width=4.35cm]{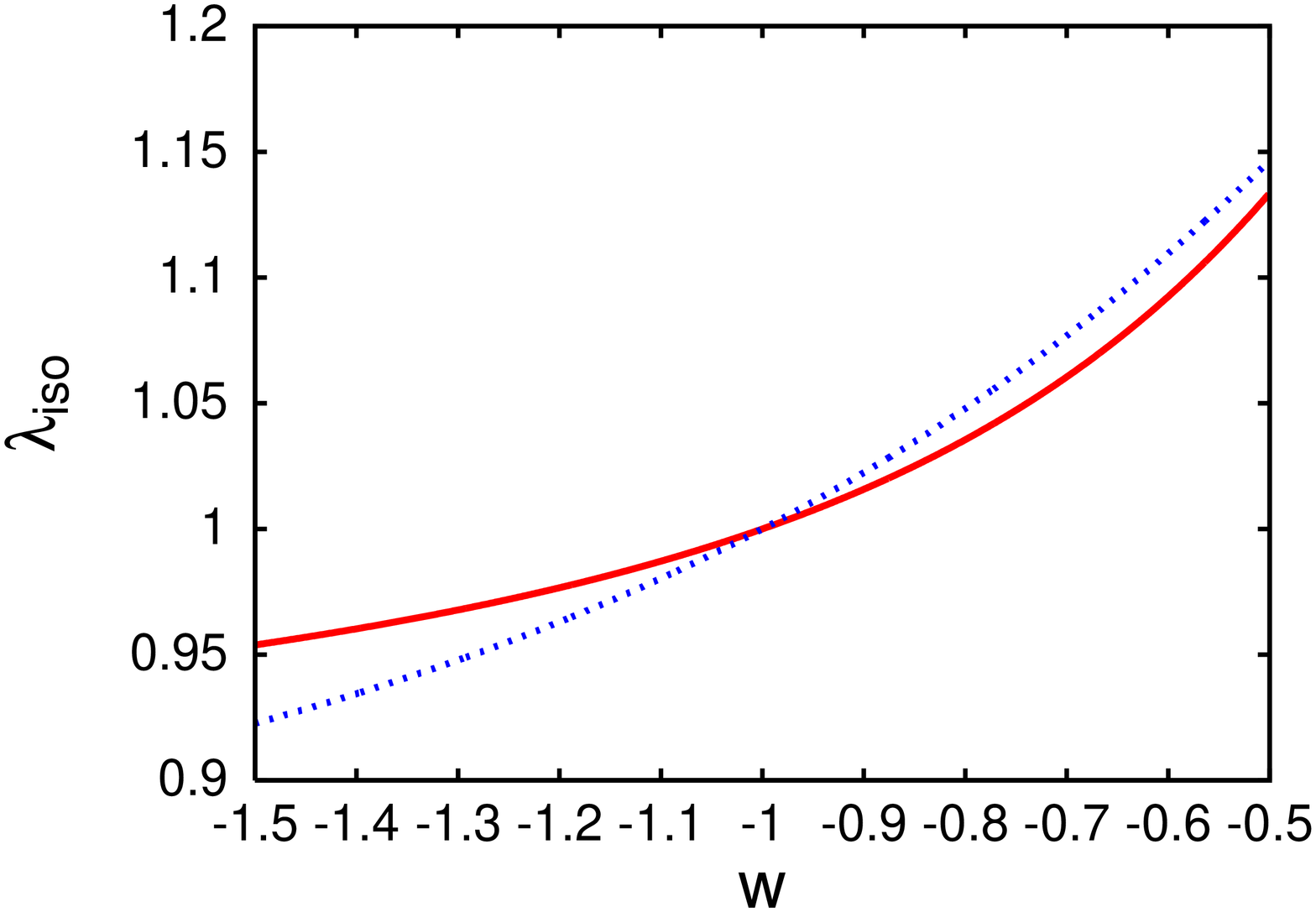}
\includegraphics[angle=-90,width=4.35cm]{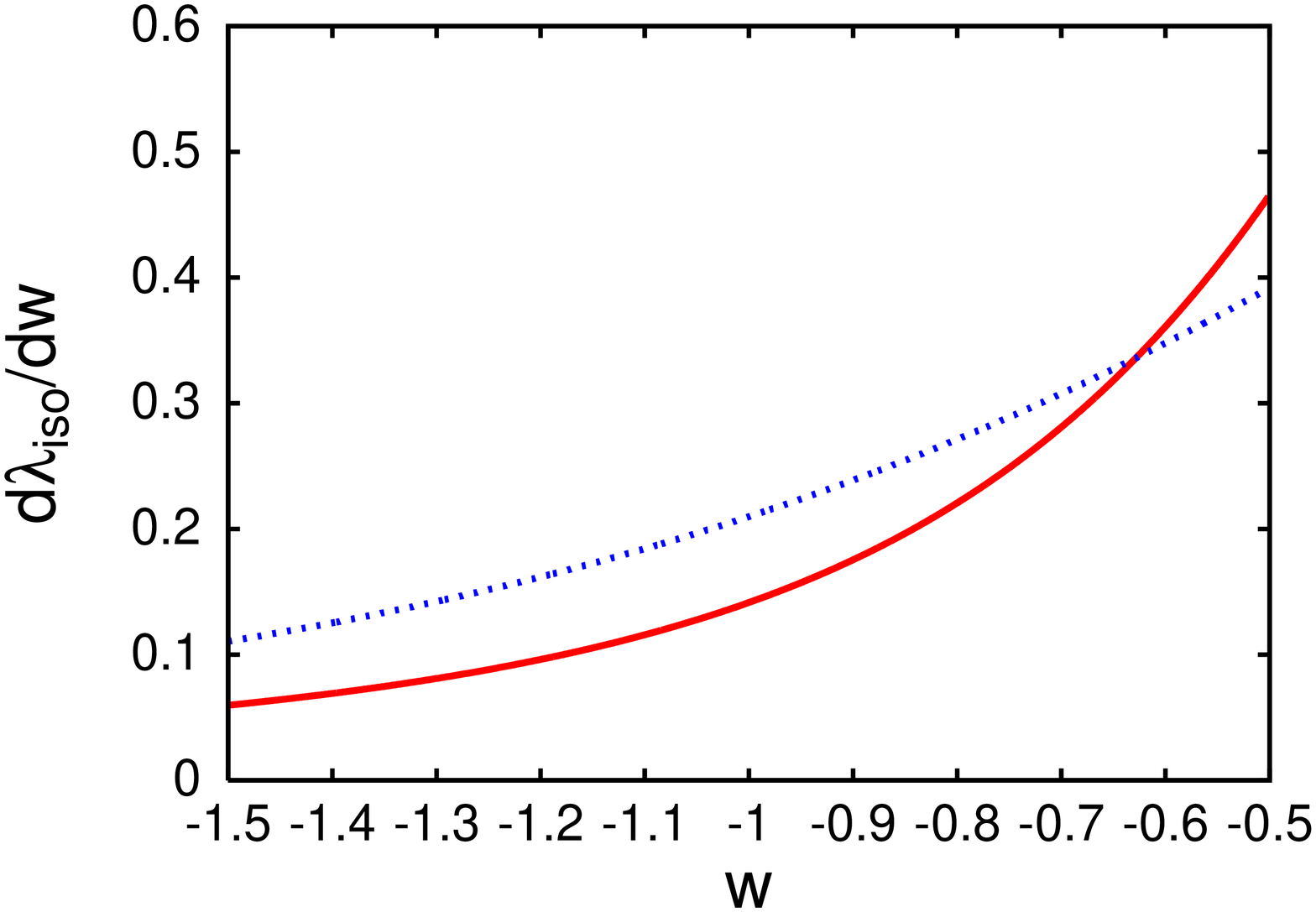}
\caption{\footnotesize The scaling factors $\lambda_{\parallel,\bot}$ and
$\lambda_{\rm iso}$ and their derivatives are shown as a function of $w$ for 
the redshifts $z=3$ (solid line)  and $z=1$ (dotted line).}
\label{fig:lambda_w}
\end{figure}

In Fig.~\ref{fig:lambda_w}, we show the dependence of the scaling factors on a 
constant $w$ for redshift $z=3$ (solid line) and $z=1$ (dotted line). As a 
reference value, we chose $w_{\rm ref}=-1.0$. We also show the derivatives of 
the scaling factors with respect to $w$. The higher the value of the
derivative, the more accurately
we can constrain the constant $w$ for a given uncertainty in the scaling
factor. Since the derivatives decrease with decreasing $w$, the uncertainty 
in $w$ will be, in general, higher towards lower $w$ values. The derivative 
of $\lambda_{\rm iso}$ around $w=-1$ at redshift $z=1$ is $\sim 1.5$ times 
higher than at $z=3$, that is although we can probably measure the scale of 
the BAO less accurately at redshift $z=1$ than at $z=3$ (due to nonlinear 
evolution), this does not need to be the case for the EOS parameter $w$.

\section{Simulation data} 
To derive realistic samples of large galaxy surveys, we first completed an N-body 
simulation in a large box ($1.5\,\mathrm{Gpc}/h$) and constructed 
the corresponding dark matter distribution on a light cone by interpolating 
between about $20$ snapshots. We calculated the redshifts, which would 
be observed including the effect of peculiar velocities. Assuming a certain 
cosmology, we converted these redshifts into distances. Finally, we selected a 
certain number of particles by applying a simple bias scheme and defining them as
galaxies. In the following subsections, we describe each step of this procedure.

\subsection{N-body simulation}
Our principal N-body simulation consists of $512^3$ dark matter particles of a mass 
of $2 \times 10^{12} M_{\sun}/h$ in a $(1.5\, \mathrm{Gpc}/h)^3$ box. The 
initial power spectrum was produced by CMBfast \citep{cmbfast}. Starting from 
a glass distribution, the particles were displaced according to second order 
Lagrangian perturbation theory by using the code of \citet{sirko}. As 
cosmological parameters, we chose $\Omega_{\rm m}=0.27$, $\Omega_\Lambda=0.73$, 
$h=0.7$, $\Omega_{\rm b} h^2=0.023$, $n_s=0.95$, $\sigma_8=0.8$ and $w=-1.0$. 
The simulation was performed with GADGET-2 \citep{gadget2} using a softening
length of comoving $100\,\mathrm{kpc}/h$. The starting redshift was $z=20$.

We also completed twelve $256^3$ dark matter particles simulations with 
the same cosmological parameters but different realizations of the initial 
conditions. We used these simulations to investigate systematic effects. 
Especially, we tested if our fitting procedure provides unbiased results.

\subsection{Light-cone survey}
An observer at the present epoch $t_0$ ($z(t_0)=0$) identifies the galaxy
distribution on his past light cone. He receives photons emitted at $t<t_0$
that have traversed the comoving distance 
$\chi(t)=c\int_t^{t_0}{a^{-1}(\tilde t)} \mathrm{d}\tilde t$. 
To construct a light-cone survey, we follow an approach used by 
\citet{lightcone}. We identify for each particle two consecutive 
snapshots between which it crosses the light cone and interpolate between 
them to find the position and velocity of that particle 
on the light cone. Expressed in formulas, the interpolated position is given by 
$\vec{x}=\vec{x_i}+\alpha \Delta \vec{x}$ where $\vec{x_i}$ is the position 
of the particle in snapshot $i$ and $\Delta\vec{x}=\vec{x_{i+1}}-\vec{x_i}$, 
and $\alpha$ is determined by requiring that $|\vec{x}|=\chi(t_i+\alpha\Delta t)$ 
with $\Delta t$ being the time step between the two snapshots. After a Taylor 
expansion for the last term, we can solve for $\alpha$
\begin{align}
\alpha=\frac{\chi^2(t_i)-x_i^2}{2 (\vec{x_i}\cdot \Delta\vec{x}+ \chi(t_i)\Delta \chi)}\,,
\end{align}
with $\Delta \chi=\chi(t_i)-\chi(t_{i+1})>0$. The interpolated velocity is 
then given by $\vec{v}=(1-\alpha)\vec{v_i}+\alpha\vec{v_{i+1}}$.

For the galaxy sample at redshift $z=3$, we place the center of the simulation
box at redshift $z=3$, 
i.e.~$6.4\, \mathrm{Gpc}$ comoving distance away from the virtual observer. 
The orientation of the box is such that the line connecting the center of the
box with the observer is parallel to the $z$-axis.
The $(1.5 {\rm Gpc}/h)^3$ box
then extends from redshift 2.1 to 4.7, i.e.~from 5.3 Gpc to 7.6 Gpc in
comoving distances. For this redshift range, we have 17 snapshots at different 
times, i.e.~of different expansion factors: $a=0.17, 0.18, \ldots 0.33$. The 
box centered at redshift $z=1$ ranges from redshift 0.6 to 1.7. For this 
interval, we use 27 snapshots with $a=0.37, 0.38, \ldots 0.63$ for the light
cone construction.

\subsection{``Observation''}
\label{observation}
Our virtual observer is at redshift $z=0$ and the simulation box 
is centered on redshift $z=3$ and $z=1$, respectively. To derive a redshift for 
each particle, we compute the comoving distance $\chi$ to the observer
and solve the following equation numerically for the expansion factor $a$, 
\begin{align}
\chi=c\int_{a}^1\frac{\mathrm{d}\tilde a}{\tilde a^2 H(\tilde a)}\,,
\end{align}
where $H$ is the ``true'' Hubble parameter of the simulation 
$H=H_0\sqrt{\Omega_{\rm m}/a^3+(1-\Omega_{\rm m})}$.
The corresponding redshift is then $z_H=1/a-1$. In addition to this redshift 
caused by the Hubble expansion of the Universe, there is also a Doppler
redshift caused by the peculiar velocity of the particle. Hence, the total 
observed redshift is $z=(1+z_H)(1+{v_\mathrm{rad}}/{c})-1$, where 
$v_\mathrm{rad}$ is the radial velocity of the particle. The accuracy of the 
simulation and the numerical algorithm of the integral equation above, implies 
an uncertainty of $\Delta z \sim 0.001$.

We next convert the observed redshifts to distances by assuming different reference cosmologies, where the dark energy EOS parameter
$w$ has the values $w_{\rm{ref}}=-0.8$, $-1.0$, and $-1.2$. In this case, the
Hubble parameter becomes 
$H^{\rm ref}(a)= H_0\sqrt{\Omega_{\rm m}/a^3+(1-\Omega_{\rm m})/a^{3(1+w_{\mathrm{ref}})}}$.

\subsection{Galaxy Bias}

Since the mass resolution of our simulation is too low for identifying galaxy-sized
 friends-of-friends halos, we use a simple bias scheme  
to compile mock galaxy samples from the dark matter distribution \citep{cole98,yoshida}.
Using the initial density field, we define a probability function by 
$P(\nu) \propto (\nu-\nu_{\rm th})^\alpha$ if $\nu$ is above the threshold 
$\nu_{\rm th}$ and zero otherwise. The dimensionless variable $\nu$ is given
by the density contrast normalized by its root-mean-square on the 
grid $\sigma$, i.e~$\nu(\vec x)=\delta(\vec x)/\sigma$. The density contrast $\delta(\vec x)$ of each particle is computed in the following way. First, we assign the 
particles with the cloud in cell (CIC) scheme to a $512^3$ grid
\citep{hockney} to obtain the density $\rho$ on the grid points. Then, we 
calculate the density contrast $\delta(\vec x)=(\rho(\vec x)-\bar\rho)/\bar \rho$ 
on the grid points and interpolate it to the positions of the particles.

In the next step, we Poisson sample the dark matter particles according to the 
probability function and track these ``galaxies''
throughout the snapshots and the constructed light cones.

\begin{figure}[htbp]
\centering
\includegraphics[angle=-90,width=9cm]{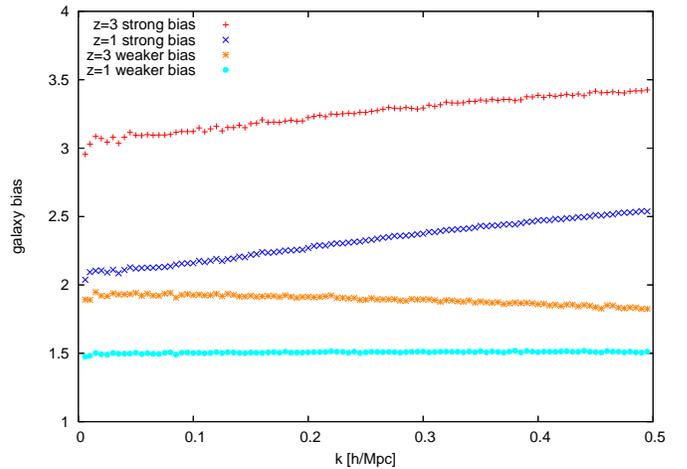}
\caption{\footnotesize Galaxy bias for the two different parameter sets at 
redshift $z=3$ and $z=1$. To construct this plot, we used 15 million galaxies to reduce 
the shot noise.}
\label{fig:bias}
\end{figure}

As parameters, we use ($\alpha=0.3$, $\nu_{\rm th}=0.4$) for a strongly 
biased sample and ($\alpha=0.2$, $\nu_{\rm th}=-0.6$) for a more weakly biased
sample. We calculate the corresponding galaxy bias to be the square root of 
the ratio of the galaxy power spectrum to the dark matter power spectrum 
in real space: $b^2=P_{\rm gal}(k)/P_{\rm dm}(k)$. 
The bias of the mock galaxies decreases with decreasing redshift (Fig.\ref{fig:bias}) 
as expected in the model. We select primarily particles in high-density
regions of the initial density field, which occupy less 
prominent structures at later times due to the further development of gravitational 
clustering. The bias has a mild scale dependence due to the schematic
procedure of grid-based density estimation and the specific probability
selection function.

The parameters $\alpha$ and $\nu_{\rm th}$ were chosen such that 
the strong bias sample at redshift $z=3$ is consistent with the expected bias 
of the target galaxies of HETDEX \citep{hetdex}. The bias of this sample is very similar to the 
friends-of-friends halo sample with a halo mass higher than 
$\sim 4 \times 10^{11} M_{\sun} / h$ obtained from the Mare Nostrum 
simulation \citep{marenostrum}. As observations of the target galaxies of 
HETDEX, namely Lyman-$\alpha$ emitting galaxies (LAE), suggest 
\citep{gronwall, gawiser}, this expected value for the bias might be too 
optimistic. Therefore, we also use the weaker bias sample at $z=3$, 
which matches the measured bias of LAE observed by the MUSYC 
collaboration \citep{gronwall, gawiser}.

For $z=1$, we use the ``weaker'' bias sample, which is 
fairly consistent with the expected bias of the target galaxies of WFMOS 
\citep{wfmos,kaos_book}. 
The bias of the strongly biased sample at $z=1$ is too high for these galaxies and we do not use this sample in the rest of the paper.

\subsection{Mock Catalogs}

For the survey centered on $z=3$, we generate several mock catalogs of one million galaxies from the strongly and weakly biased 
light cones in redshift space using the entire $1.5\, \rm Gpc/h$ box, which has a volume of $10\,{\rm  Gpc}^3$. 
These numbers are at the upper limit of the current baseline of the HETDEX project.

For the mock catalogs at $z=1$, we use only the galaxies on the light cone 
around $z=1$ in redshift space with a bias of $b\approx 1.5$. As a number 
of tracers, we choose one and two million galaxies in the entire box.

\section{Analyzing the data} 
We analyze the mock catalogs first by calculating the power spectrum and then 
by fitting the extracted BAO. In the following two subsections, we describe how 
we calculate the power spectrum and provide details about the fitting procedure.

\subsection{Power spectrum}
Since we use an FFT to complete the Fourier transformation, we first have to assign 
the particles to a regular grid. We consider a $1024^3$ regular grid and
select the CIC scheme to complete the mass assignment. After converting the density field 
$\rho(\vec x)$ to the density contrast 
$\delta(\vec x)=(\rho(\vec x)-\bar\rho)/\bar \rho$ and 
performing the FFT to evaluate the Fourier transform (FT) $\delta(\vec k)$, we 
compute the raw isotropic (one-dimensional) and anisotropic (two-dimensional) 
power spectrum by averaging $|\delta(\vec k)|^2$ over spherical shells 
$P^\textrm{iso.}_{\rm{raw}}(k)=\left\langle 
|\delta(\vec k)|^2\right\rangle_\textrm{shell} $ and rings 
$P^\textrm{aniso.}_{\rm{raw}}(k_\parallel,k_\bot)=\left\langle 
|\delta(\vec k)|^2\right\rangle_\textrm{ring} $, respectively. These raw 
power spectra are related to the true power spectrum in the following way \citep{jing}:
\begin{align}
P^\textrm{iso./aniso.}_{\rm{raw}}&=\left\langle \sum_{\vec m \in
    \mathbf{Z}^3 } |W(\vec k + 2 k_{\rm Ny} \vec m)|^2 P(\vec k +2 k_{\rm Ny} 
    \vec m) \right. \nonumber\\
    &\left. + \frac{1}{n}\sum_{\vec m \in \mathbf{Z}^3 } |W(\vec k + 2 k_{\rm Ny} 
    \vec m)|^2 \right\rangle_\textrm{shell/ring}\ ,
\end{align}
where $W(\vec k)$ is the FT of the mass assignment function, $k_{\rm{Ny}}$ the 
Nyquist frequency, and $n$ the number density of tracers for the density
field. The first term in the equation above reflects the convolution of the 
density field with the mass assignment function. The second term corresponds
to the shot noise. The summation over all integer vectors $\vec m$ takes 
care of the aliasing. In the case of the CIC scheme, $W(\vec k)$ takes the 
following form
\begin{align}
W(\vec k)=\left[ \prod_i {\frac{ \sin\left( \frac{\pi k_i}{2k_{\rm Ny}}
      \right) }{ \frac{\pi k_i}{2k_{\rm Ny}} }}\right] ^2\,.
\end{align}

To derive a  good estimate of the true power spectrum from its raw form, 
we first subtract the shot noise and then follow an iterative method 
for the correction of deconvolution and aliasing as proposed by \citet{jing}. 
We subtract only for the galaxy samples shot noise, since 
we do not detect any shot noise for the dark matter densities.

We estimate the error in the power spectrum 
by counting the number of independent modes used in the calculation 
of $P_{\rm raw}$, which corresponds approximately to 
$\sigma^\textrm{ring}_P=
\sqrt{\frac{4 \pi^2}{k^2\Delta k \Delta \mu V_{\rm survey}}}
\left(P+N_s\right)$ and 
$\sigma^\textrm{shell}_P=
\sqrt{\frac{4 \pi^2}{k^2\Delta k  V_{\rm survey}}}
\left(P+N_s\right)$ \citep{feldman} 
for the anisotropic and isotropic power spectrum, respectively. Here, $\Delta k$ 
denotes the thickness of the rings and shells we averaged over. 
For the anisotropic power spectrum, there is additionally the parameter
$\Delta \mu$ which is the range in the values of the cosine of the angle between the wave vector and the line 
of sight. The shot noise is denoted by $N_s$. All the power 
spectra in this paper have independent bins with a bin width of 
$\Delta k=0.005\, h/\rm{Mpc}$.

\begin{figure}[htbp]
\centering
\includegraphics[angle=-90,width=9cm]{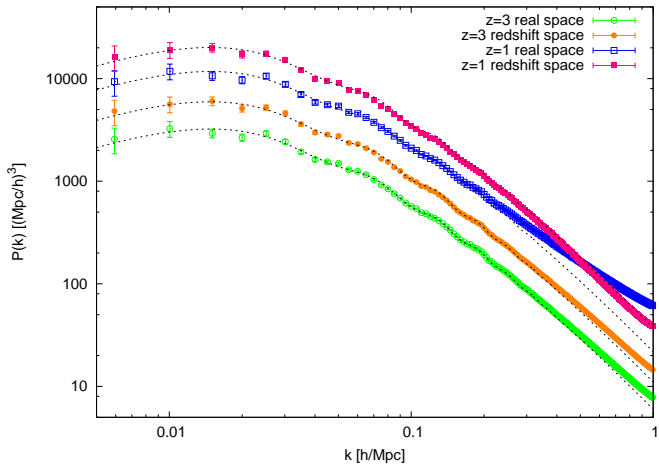}
\caption{\footnotesize Power spectrum of the dark matter distribution at $z=3$ 
(lower curves) and $z=1$ (upper curves). The dashed lines are the linearly 
evolved initial power spectrum. In the case of redshift space, 
the linear power spectrum was multiplied in addition by $1+2/3\, 
\beta(z) + 1/5\,\beta(z)^2\approx 1.843 \, (1.714)$ for $z=3\,(1)$.}
\label{fig:ps_w10}
\end{figure}

\begin{figure}[htbp]
\centering
\includegraphics[angle=-90,width=9cm]{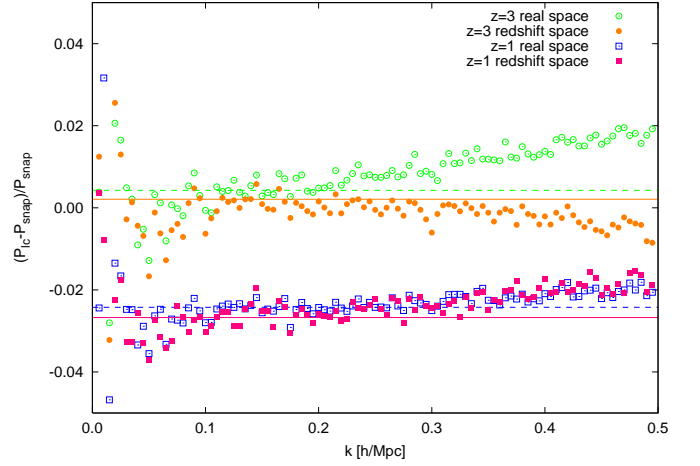}
\caption{\footnotesize Fractional difference of the light-cone power spectrum 
$P_{\rm lc}$ and the power spectrum at the corresponding snapshot 
$P_{\rm  snap}$. The horizontal lines show the predicted shifts from linear 
theory in real space (dashed) and redshift space (solid).}
\label{fig:ps_ratio}
\end{figure}

\begin{figure}[htbp]
\centering
\includegraphics[angle=-90,width=9cm]{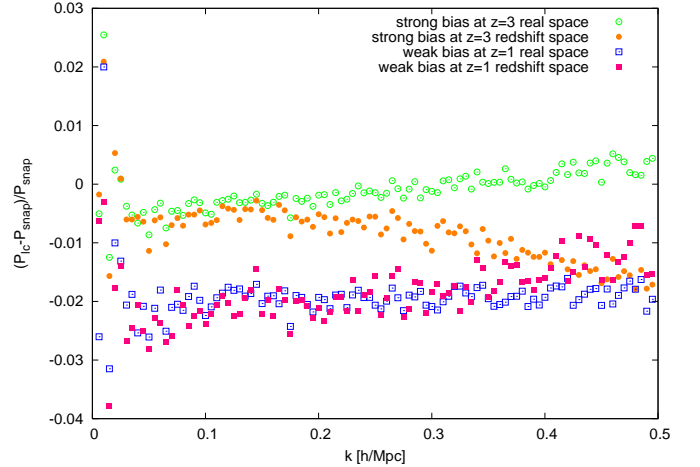}
\caption{\footnotesize Fractional difference of the galaxy light-cone power 
spectrum $P_{\rm lc}$ and the galaxy power spectrum at the corresponding snapshot 
$P_{\rm  snap}$.}
\label{fig:ps_ratio_bias}
\end{figure}

In Fig.~\ref{fig:ps_w10}, we show the isotropic dark matter power spectra at 
redshift $z=3$ and $z=1$ obtained by assuming the correct cosmology with $w_{\rm
  ref}=-1$. 
If not stated otherwise, we always use as the reference cosmology the cosmology of 
the simulation, i.e.~we assume that $w_{\rm ref}=-1$. The linear redshift distortion 
\citep{kaiser} amplifies the power by $1+2/3\, \beta(z) + 1/5\,\beta(z)^2$ 
with $\beta(z)=f(z)/b(z)$, where $f(z)$ is the linear growth rate and $b(z)$ 
the bias parameter. In the case of dark matter we have $b=1$. 
For scales shown, the nonlinear evolution is still mild. The 
deviations in the real-space power spectra from the linearly evolved initial 
power spectra (plotted as dashed lines) start around $k\sim 0.25 h/{\rm Mpc}$ 
and $k\sim 0.15 h/{\rm Mpc}$ for $z=3$  and $z=1$, respectively. The 
nonlinear redshift distortions (``finger of God'' effects) in redshift space 
lead to a suppression of power on small scales, which almost balances  
the increase in power due to nonlinear clustering at $z=1$ (filled squares in Fig.~\ref{fig:ps_w10}). 
The BAO can be seen as tiny wiggles in 
the power spectrum. Their amplitudes are $\lesssim 5\%$ of $P(k)$. 

We do not show the light-cone power spectra, since they lie almost exactly 
on the corresponding snapshot power spectra. The fractional differences in
the power spectra derived from the light cone and the corresponding snapshot  
are instead plotted in Fig. \ref{fig:ps_ratio}. On linear scales, the fractional 
differences are $\sim 0.5\%$ and $\sim -2.5\%$ at $z=3$ and $z=1$, respectively. 
The reason for these differences is that the value of the growth function at 
the mean redshift is not equal to the mean growth function.
The stated numbers can be understood by averaging the square of the 
redshift-dependent growth function multiplied by an appropriate geometrical 
factor over the survey box \citep{ratio}. 
Numerical calculations for our survey designs are shown as dashed (real space) 
and solid (redshift space) lines. The deviations at larger $k$ are due 
to nonlinear effects. The question of whether these light-cone effects alter 
the fitting of the BAO is addressed in Sect.~\ref{results}.

In Fig. \ref{fig:ps_ratio_bias}, we compare the biased galaxy power spectra 
derived from the light cones with those evaluated for the corresponding 
snapshots. Qualitatively, we observe the same behavior as in the dark matter 
case. The larger scatter is due to higher shot noise and the means are 
shifted slightly, since, as happened before for the growth function, the bias at 
the mean redshift is not equal to the bias averaged over the appropriate 
redshift range.

\subsection{Fitting procedure}

\begin{figure*}[!htbp]
\centering
\includegraphics[angle=-90,width=8.8cm]{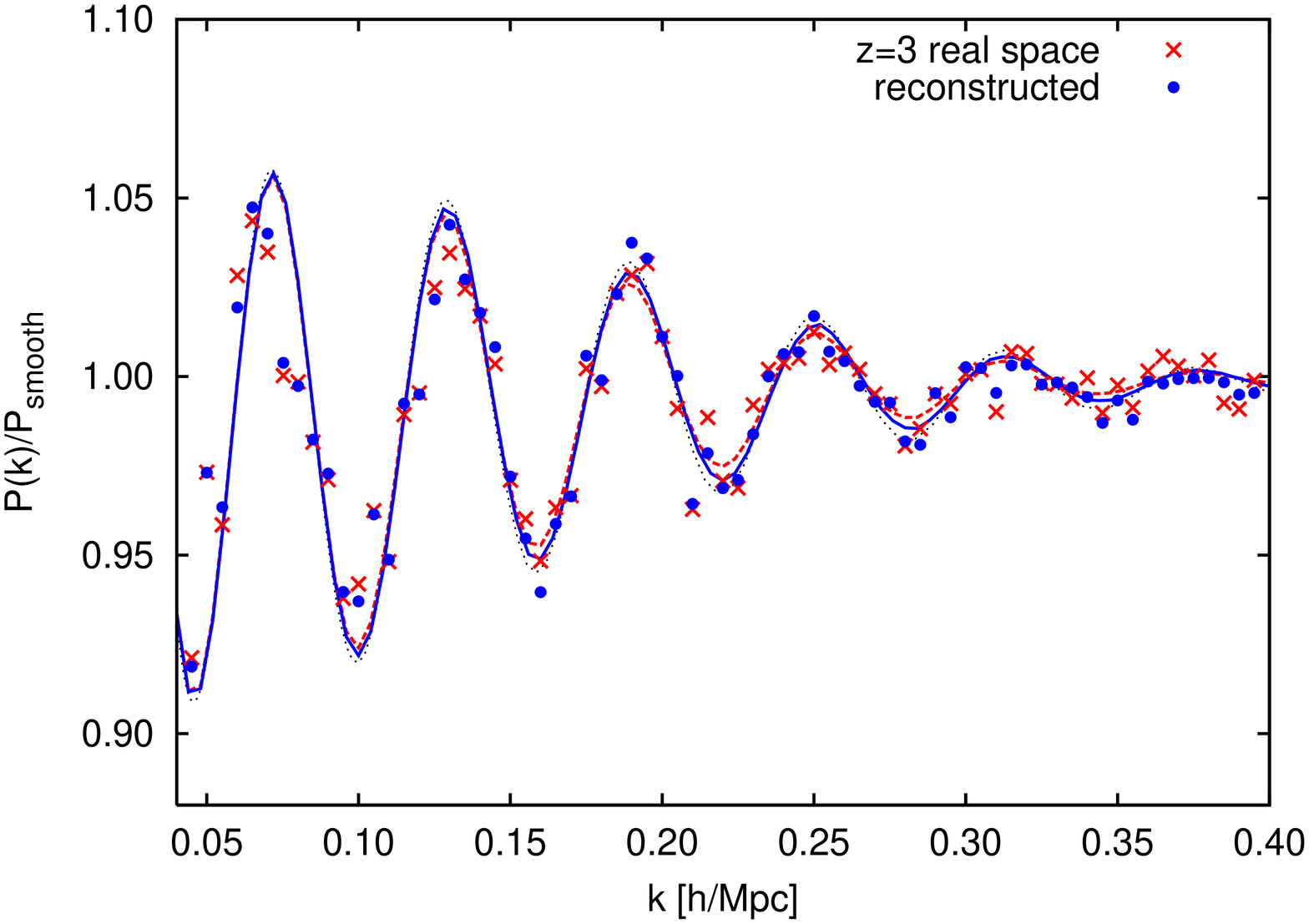}
\includegraphics[angle=-90,width=8.8cm]{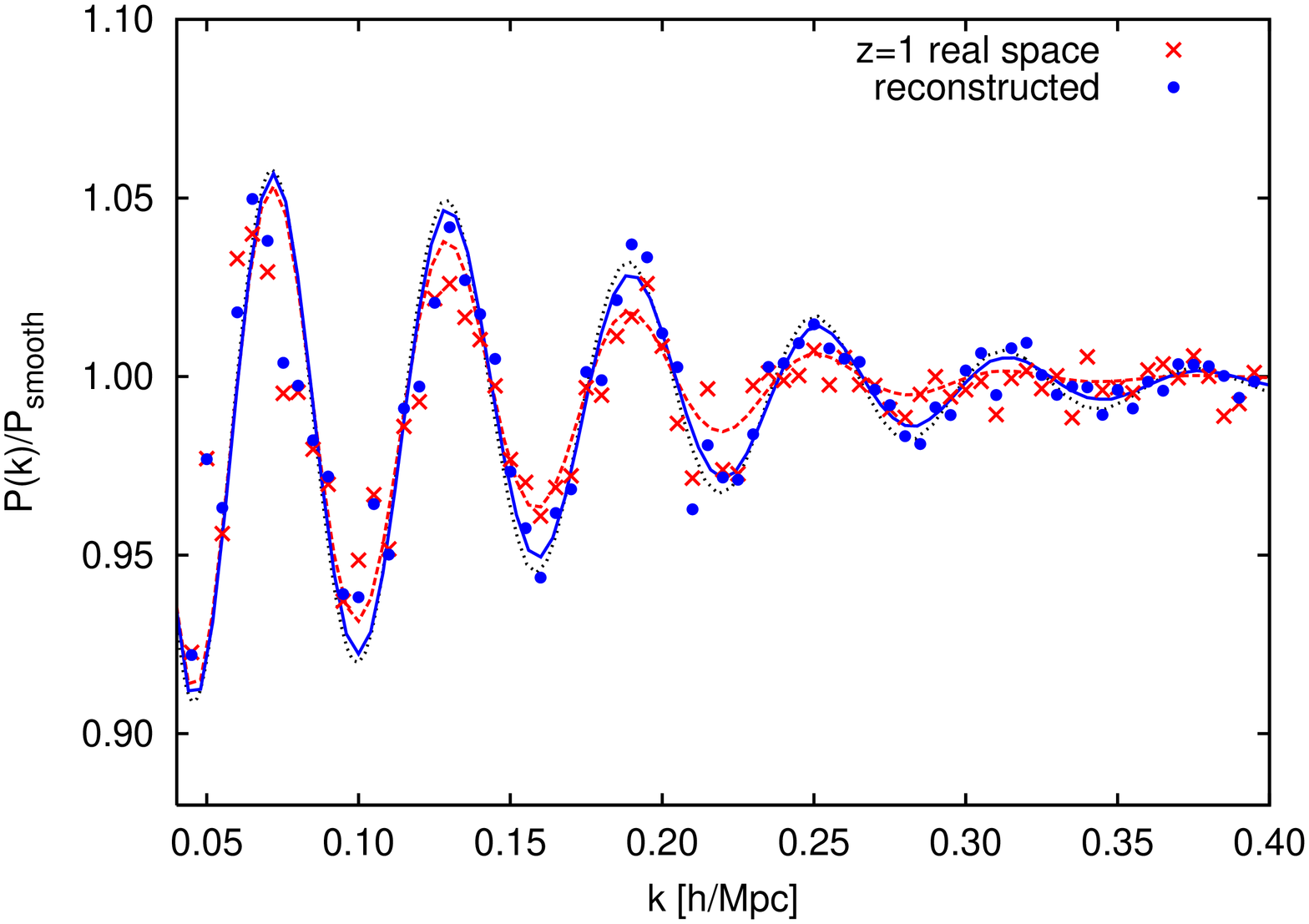}
\includegraphics[angle=-90,width=8.8cm]{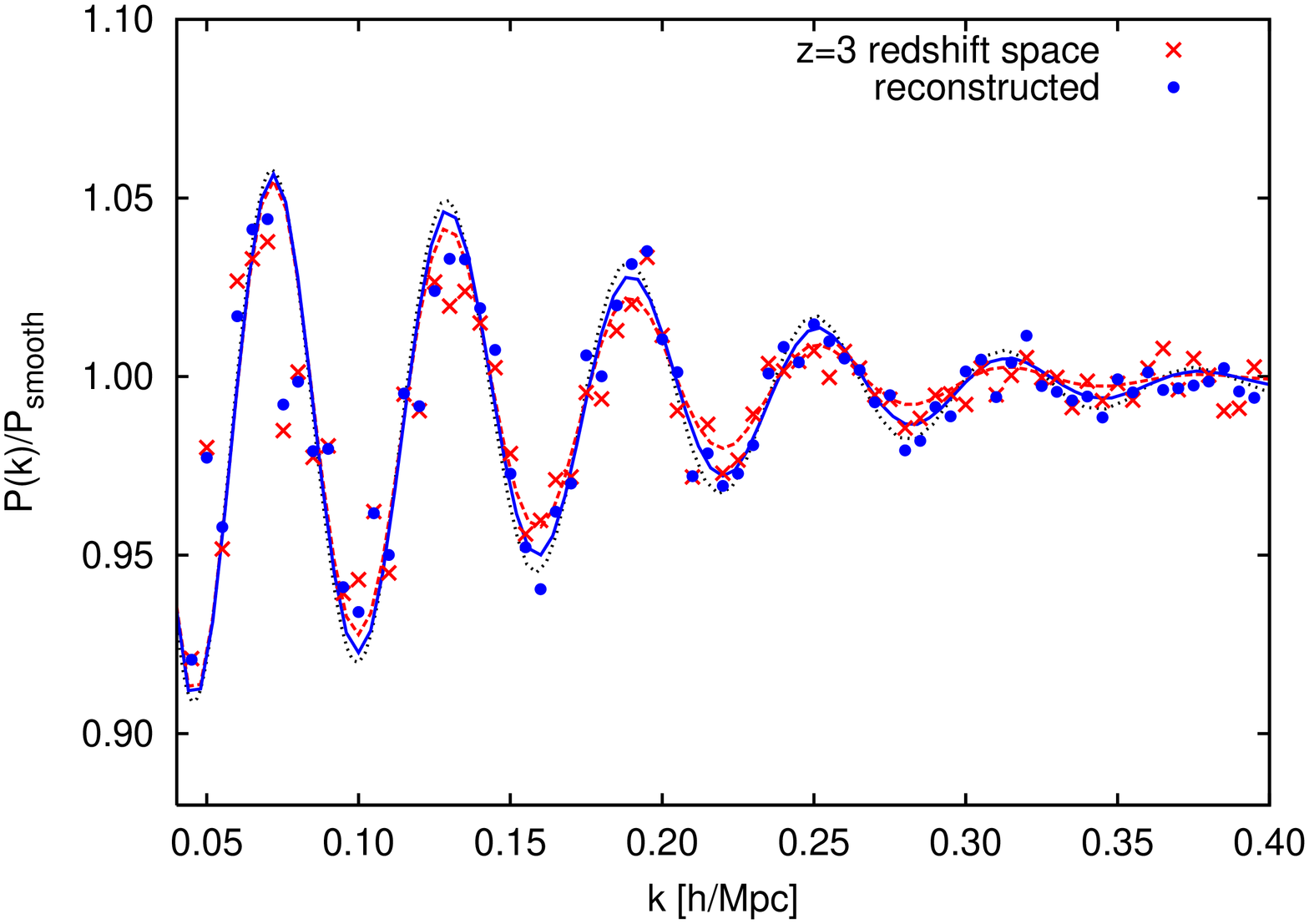}
\includegraphics[angle=-90,width=8.8cm]{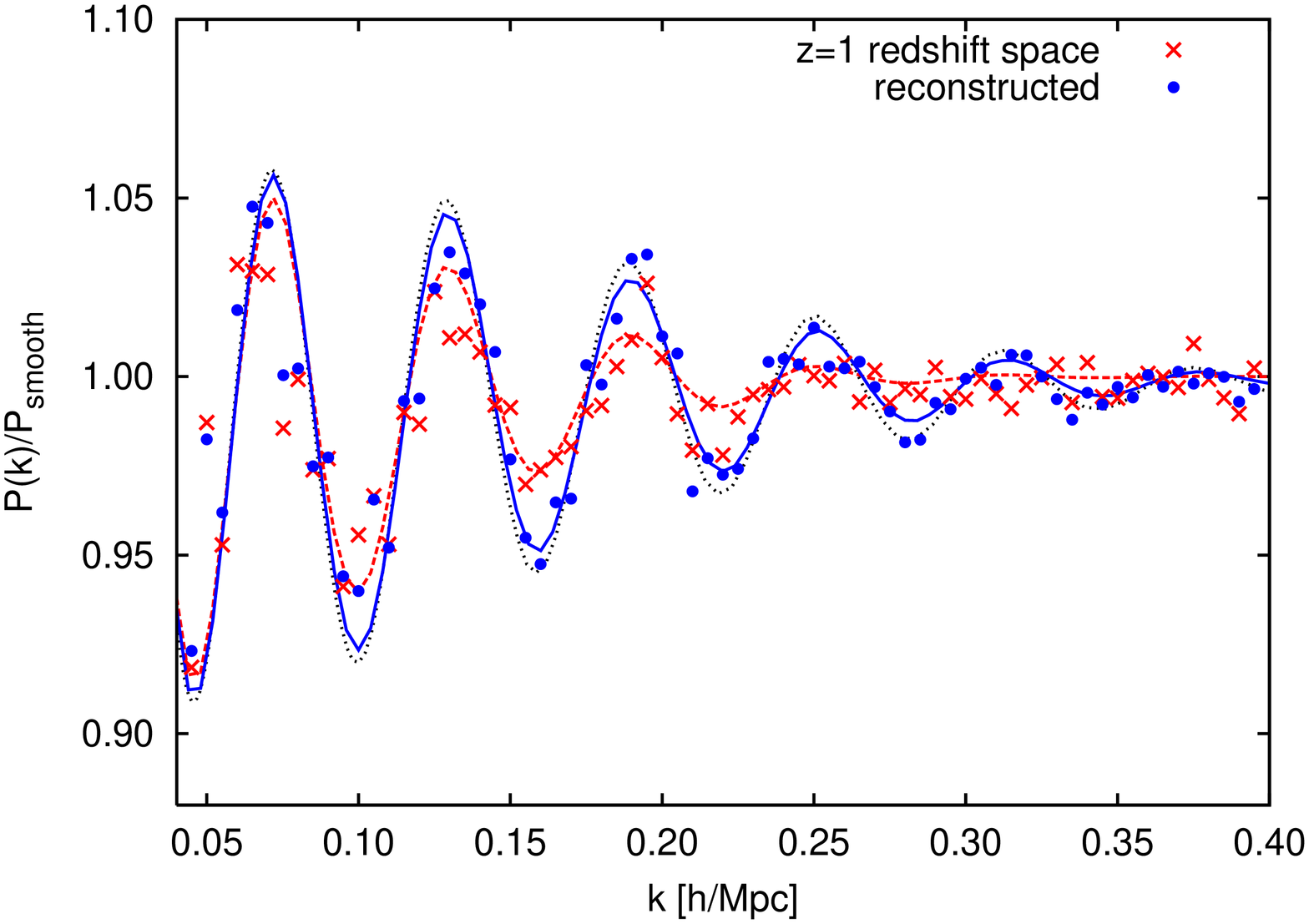}

\caption{\footnotesize The left upper panel shows the oscillatory part of the 
power spectrum at redshift $z=3$ obtained from real space and its best-fit
function including the suppression factor (dashed line). The dotted line
corresponds to the oscillatory part of the initial power spectrum. The solid 
line is the best-fit of the reconstructed data.
At the bottom left, we present the same in redshift space. On the right the 
corresponding plots for $z=1$ are shown.}
\label{fig:osci}
\end{figure*}

Our fitting method attempts to remove all the information 
apart from the BAO themselves. This is easily achieved by dividing the power 
spectrum by a smoothed version of itself. The advantage of this is that one does
not need to model all the physical processes that alter the power spectrum
such as redshift distortion, nonlinear evolution, and galaxy bias or uncertainties 
in cosmological parameters, such as $\sigma_8$, $n_s$, and massive neutrinos,
which affect the overall shape of the power spectrum but have little effect 
on the BAO. 

The extracted BAO of the reference power spectra were then fitted to the
extracted BAO, allowing the scaling factors $\lambda_{\parallel,\bot}$ or 
$\lambda_{\rm iso}$ to vary.
In an additional fitting attempt, our free parameter is instead the EOS parameter $w$ and we derive the scaling factors $\lambda_{\parallel,\bot}$ or 
$\lambda_{\rm iso}$ by applying Eq. (\ref{scalingfactors}).

The fitting parameters are determined by Monte Carlo Markov
chain (MCMC) techniques. We use the Metropolis-Hastings algorithm 
\citep{metropolis,hastings} to build up the Markov chain.

In the following, we describe each step of the fitting procedure.

\subsubsection*{Smoothing}
There are previous proposals to use only the oscillatory part of the power
spectrum for measuring the scale of the BAO 
\citep{blake_2003,blake_2,huetsi_astro_ph,huetsi_sdss,koehler,angulo2007,percival}. 
All of them divide the power spectrum by (or subtract from it) a non-oscillating 
fit. The methods used to derive the smooth fit range from deriving a semi-analytic 
zero-baryon reference power spectrum, fitting with a quadratic polynomial 
in log-log space, or using a cubic spline, to fitting with a non-oscillating 
phenomenological function. In this paper, we generate a smoothed power spectrum in an
almost \textit{parameter-free} way, by computing for each point the arithmetic mean 
in log space of its neighbors in a range of $\pm 0.03 \,h/\rm{Mpc}$ in $k$. In the 
two-dimensional case, we smooth radially in the direction of $|k|$. This
smoothing length is the only parameter in our smoothing method, which, for all input power spectra, we select to have the same value of approximately equal to the wavelength of the BAO.

By dividing the measured power spectrum by its smoothed version, we derive the 
purely oscillatory part of the power spectrum 
$P_{\rm{osci}} = P / P_{\rm{smooth}}$ (see Fig. \ref{fig:osci}).

\subsubsection*{Fitting and Priors}

In place of fitting the extracted BAO by a (modified) sine function
\citep{blake_2003,blake_2} 
or using a periodogram \citep{huetsi_astro_ph}
to measure the scale of the BAO, we compare the BAO with a range of different oscillatory reference
power spectra $P_{\rm osci}^{\rm ref}$ produced from thousands of linear
power spectra generated with CMBfast, which differ in the cosmological
parameters $\Omega_{\rm m}$, $\Omega_{\rm b}$, $H_{\rm 0}$, and $n_{\rm s}$. 

We prefer this method since the BAO are not exactly harmonic
\citep{ehu98,koehler} and the uncertainties in the aforementioned
cosmological parameters can easily be included. As priors on these
cosmological parameters we use the predicted uncertainties for the 
Planck mission \citep{planck}. 
For the Hubble constant $H_0$, we combine the Planck priors with constraints
provided by large-scale structure \citep{bao_z0.3} and include measurements from the HST Key project \citep{HST}: 
\begin{align}
\begin{aligned}
\Delta \omega_{\rm m}&= 1.25\%\ , \ \ \Delta H_{\rm 0}= 5\%\ ,\\
\Delta \omega_{\rm b}&= 0.75\%\ , \ \ \Delta n_{\rm s}= 0.5\%\ ,
\end{aligned}
\end{align}
where $\omega_{\rm m}=\Omega_{\rm m}\,h^2$ and $\omega_{\rm b}=\Omega_{\rm b}\,h^2$.
We assume throughout this article that the Universe is spatially flat.

We determine the scaling factors 
$\lambda_{\parallel,\bot}$ or $\lambda_{\rm iso}$ by fitting a scaled 
$P_{\rm osci}^{\rm ref}(\lambda_{\parallel} k_{\parallel}, 
\lambda_{\bot} k_{\bot})$ or $P_{\rm osci}^{\rm ref}(\lambda_{\rm iso} k)$ 
to $P_{\rm{osci}}$.

\subsubsection*{Damping and Reconstruction}

Since nonlinear structure growth diminishes the amplitudes of the wiggles 
\citep{unwiggle}, we can improve the fitting by adding a suppression factor 
$W=\exp(-\sigma k^2)$ to our fitting function
\begin{equation}
P_{\rm osci}(k)=W \left[P_{\rm osci}^{\rm ref}(\lambda_{\rm iso} k) - 1\right] +1  \,.
\end{equation}
For the two-dimensional redshift-space power spectrum, we have to introduce 
two suppression parameters $\sigma_{\parallel,\bot}$, since the redshift space
distortion increases suppression along the line of sight. The suppression
factor then becomes $W=\exp(-\sigma_\parallel k_\parallel^2-\sigma_\bot k_\bot^2)$.

If the density field is known accurately, i.e.~the shot noise is small and the
galaxy bias is known, it is possible to use reconstruction techniques 
(see e.g. \citet{linrecon} for a comparison of methods) to undo, at least
in part, the suppression of the wiggles \citep{eisensteinrec}. We applied the PIZA 
method \citep{piza} to the dark matter distributions. We attempted a similar 
technique for the mock catalogs but due to shot noise and galaxy 
bias our results were unsatisfactory. Further efforts are required to achieve 
this goal. In Fig. \ref{fig:osci}, we indicate the BAO extracted from the data by 
crosses. The dashed lines show the best-fit functions and the dotted lines depict 
the corresponding non-damped reference $P_{\rm osci}^{\rm ref}(k)$. 
We observe that the suppression of the BAO is higher in redshift space and
increases with time. Additionally, the results of the reconstructed density 
fields are plotted (data as filled circles and the best-fits as solid lines). 
The amplitude of the BAO could not be reestablished completely but to a 
significant fraction.

\subsubsection*{Fitting Range}

The $k$ range over which we use the data for our fitting plays an important role. 
In the fitting procedure we use a minimum wave number $k_{\rm min}=0.04\,
h/{\rm Mpc}$, since for the points with $k<0.03\, h/{\rm Mpc}$ the smoothing
is not well defined, and the error due to sample variance is high. As the 
maximum wave number, we select $k_{\rm max}=0.25\, (0.30)$, for redshift $z=1\, (3)$. 
Although nonlinear evolution has already started on these scales, the BAO 
are still clearly visible (see Fig. \ref{fig:osci}).

\section{Results}\label{results}

The most important measurements in our analysis are the marginalized probability 
distribution functions (PDF) of the scaling factors and the EOS parameter $w$. 
To determine their values, we marginalize the joint PDF produced via the MCMC 
technique over all other fitting parameters. From these functions, we can derive 
the best-fit values and the accuracy of both the scaling factors and the EOS $w$.

\subsection*{Fitting method, real space versus redshift space, and degeneracies}

With the twelve $256^3$ simulations we assess the robustness of our fitting 
method and the presence of systematic effects. We determine the best-fitting
scaling factors, both parallel and transverse to the line of sight, for the dark matter power
spectrum in real and redshift space for all twelve simulations at redshift 
$z=10$, $3$, and $1$. For all redshifts, we find that the mean value 
of the twelve simulations is clearly less than $1\sigma$ (standard deviation) 
away from unity. Hence, our fitting method provides unbiased results within the 
margins of error (see Table \ref{tab:sf}).

\begin{table}[!htb]
\caption{\footnotesize Results for the scaling factors $\lambda_\parallel$ and 
$\lambda_\bot$ obtained from the dozen $256^3$ simulations.}
\label{tab:sf}
\centering
\begin{tabular}{c|cc|cc|cc}
\hline
\hline
 & \multicolumn{2}{c|}{Real Space} & \multicolumn{2}{c|}{Redshift Space} & 
\multicolumn{2}{c}{Difference}\\
Parameter & \multicolumn{1}{c}{mean}   & \multicolumn{1}{c|}{sigma} & 
\multicolumn{1}{c}{mean}  & \multicolumn{1}{c|}{sigma}   &
 \multicolumn{1}{c}{mean} & \multicolumn{1}{c}{sigma}\\
\hline
$z=10$ & & & & & & \\
$\lambda_{||}$  &  $0.998$ & $0.007$  &  $0.998$ & $0.008$ &  $\phantom{-}0.000$ & $0.001$ \\
$\lambda_{\bot}$  &  $0.999$ & $0.005$  &  $0.999$ & $0.005$ &  $\phantom{-}0.000$ & $0.001$ \\
\hline
$z=3$ & & & & & & \\
$\lambda_{||}$  &  $0.998$ & $0.009$  &  $0.999$ & $0.011$ &  $-0.002$ & $0.004$ \\
$\lambda_{\bot}$  &  $1.000$ & $0.006$  &  $0.999$ & $0.006$ &  $\phantom{-}0.001$ & $0.002$ \\
\hline
$z=1$ & & & & & & \\
$\lambda_{||}$  &  $0.997$ & $0.010$  &  $0.999$ & $0.014$ &  $-0.003$ & $0.008$ \\
$\lambda_{\bot}$  &  $0.999$ & $0.009$  &  $0.999$ & $0.010$ &  $\phantom{-}0.000$ & $0.003$ \\
\hline
\end{tabular}

\end{table}

The differences between results for real and redshift space in each simulation
do not show a systematic trend. The mean difference is in fact consistent with zero. The
error in the scale factor parallel to the line of sight $\lambda_{||}$ is
larger in redshift space, since the damping of the BAO is stronger in redshift 
space (see Fig \ref{fig:osci}).

\begin{figure}[htbp]
\centering
\includegraphics[width=4.4cm]{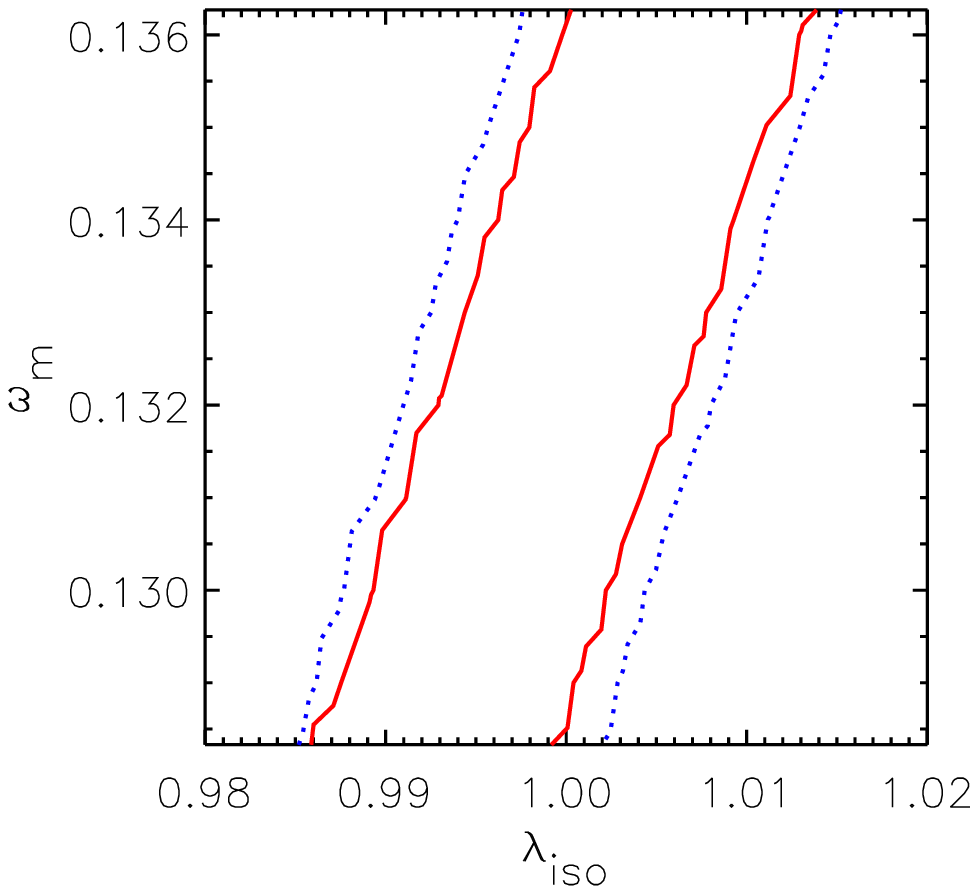}
\includegraphics[width=4.4cm]{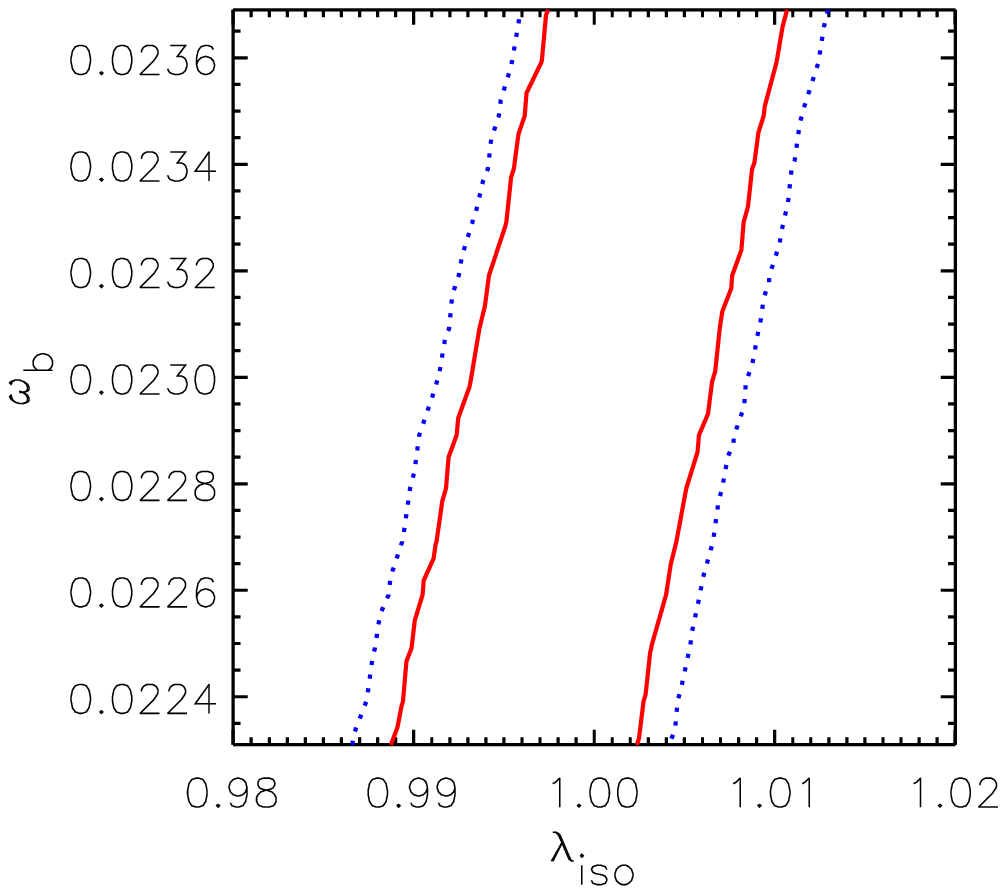}
\includegraphics[width=4.4cm]{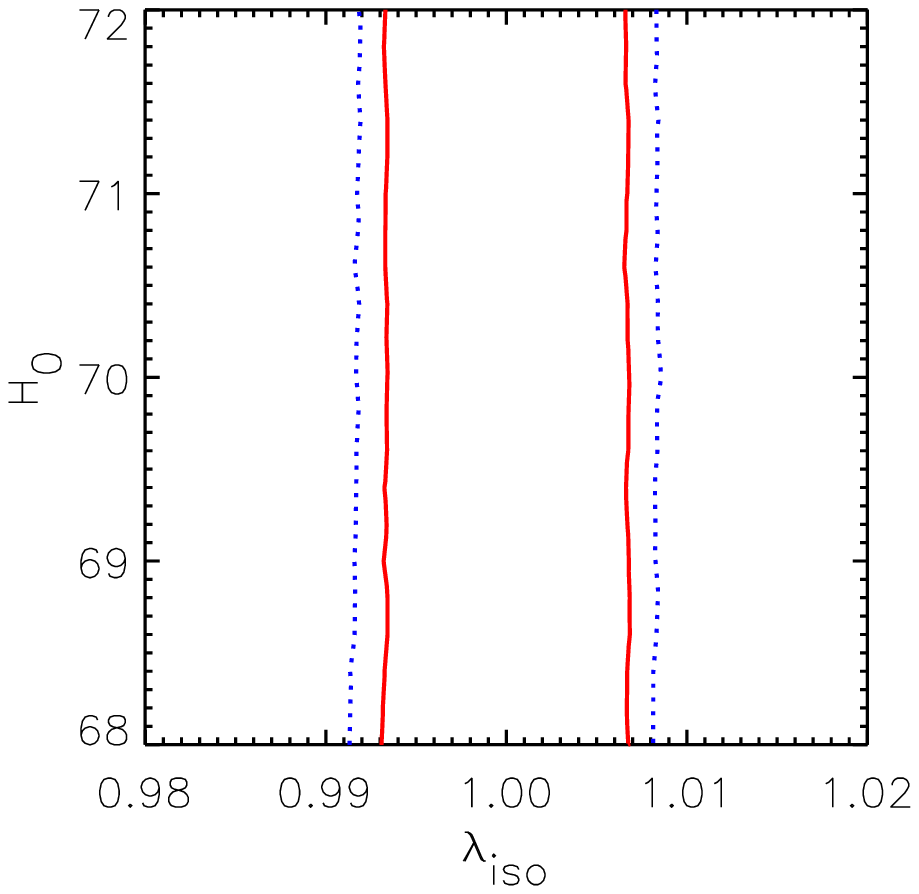}
\includegraphics[width=4.4cm]{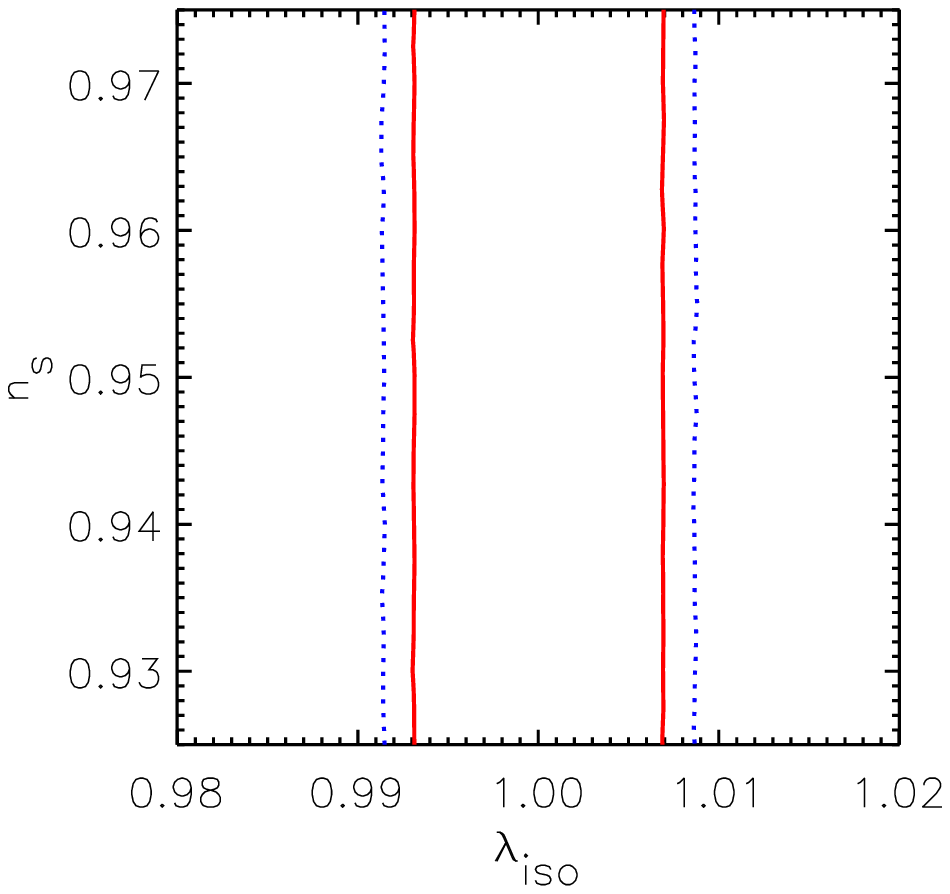}
\caption{\footnotesize Dependence of the fitted $\lambda_{\rm iso}$ on the
  cosmological parameters $\omega_{\rm m}$, $\omega_{\rm b}$, $H_0$, and $n_{\rm s}$
  at redshift $z=3$ (solid) and $z=1$ (dotted).}
\label{fig:deg}
\end{figure}

To understand the degeneracies in the scaling factor for the
cosmological parameters included in our method, we complete the fitting 
 by keeping all cosmological parameters fixed to the value of the simulation 
apart from the parameter under consideration, for which we calculate 
the PDF of $\lambda_{\rm iso}$ for different values of this parameter. 
The corresponding $95\%$ confidence lines for the dark matter power 
spectrum at $z=3$ (solid) and at $z=1$ (dotted) are shown in 
Fig. \ref{fig:deg}.\footnote{Since we used only one realization 
(the $512^3$ simulation) to produce these plots, the fitting results were
slightly off the input values of the simulation. For display purposes, we
centered the lines.} 
Since the sound horizon imprinted in the power spectrum is redshift
independent, we expect that the derived degeneracies are also
redshift independent. This is indeed the case for our fitting method. 
The only difference with respect to $z$ is the larger uncertainty in 
$\lambda_{\rm iso}$ at lower redshifts due to the suppression of the 
BAO by nonlinear evolution.

The interval of cosmological parameters plotted in Fig. \ref{fig:deg} 
was chosen to be centered on the values of the simulation  and in the range of
$\pm 3\%$. In this way, we can immediately assess the degree of degeneracy. 
We find that if we alter one of the cosmological parameters by $1\%$ 
the scaling factor changes by $0.22\%$ for $\omega_{\rm m}$, $0.14\%$ for 
$\omega_{\rm b}$, $\sim 0\%$ for $H_0$, and $0\%$ for $n_{\rm s}$. 
These numbers agree well with those given by the fitting 
formula for the sound horizon by \citet{ehu98}:
\begin{equation}
 s=\frac{44.5 \,{\ln}(9.83/\omega_{\rm m})}{\sqrt{1+10\,\omega_{\rm b}^{3/4}}}
 \,\rm Mpc \,.
\end{equation}

\begin{figure}[htbp]
\centering
\includegraphics[width=4.4cm]{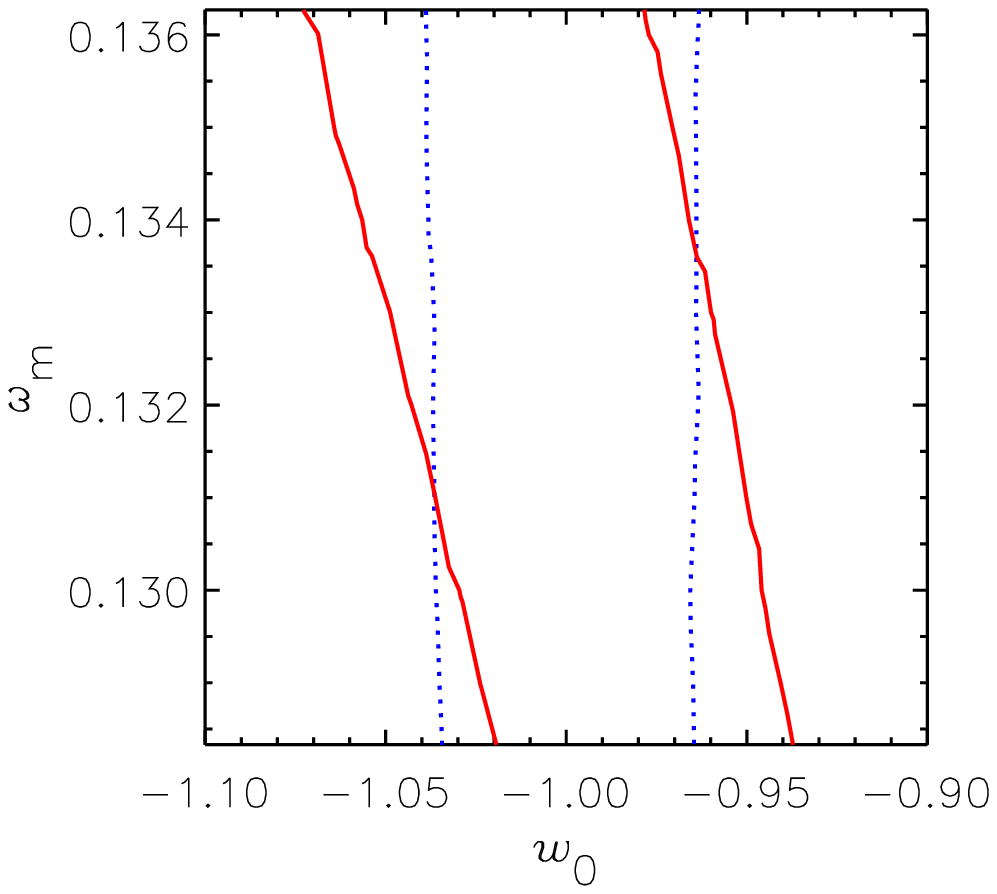}
\includegraphics[width=4.4cm]{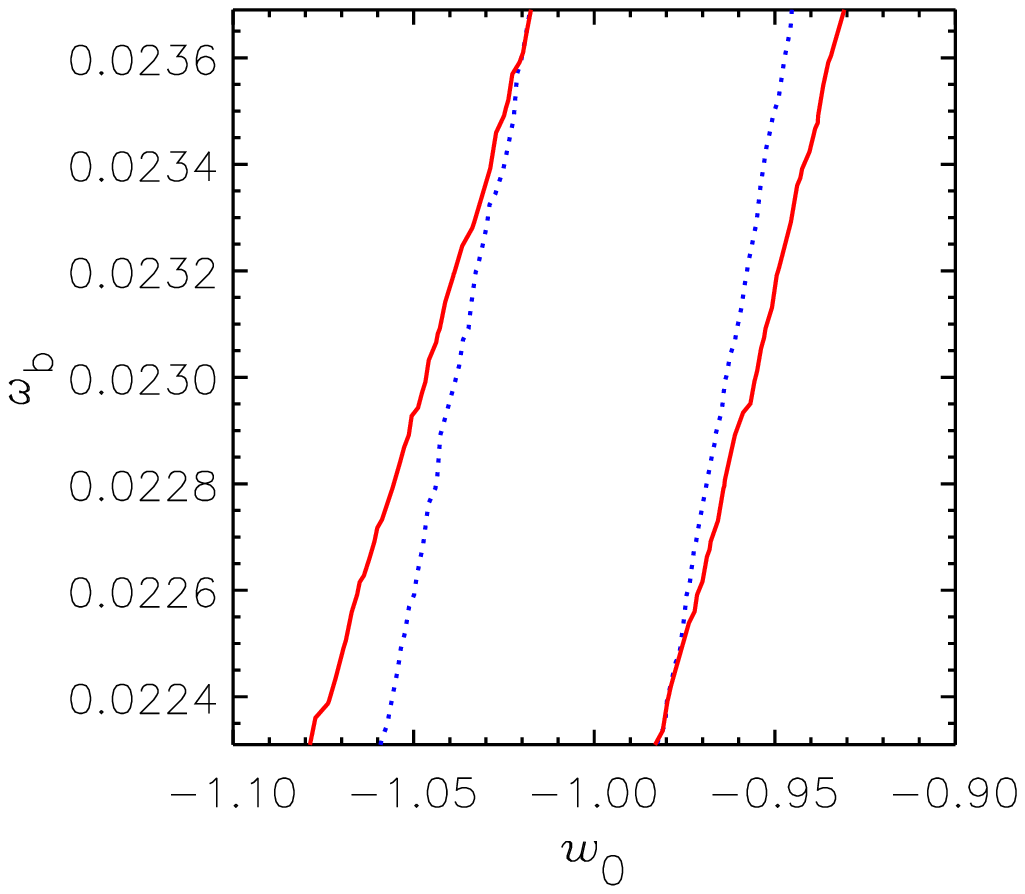}
\includegraphics[width=4.4cm]{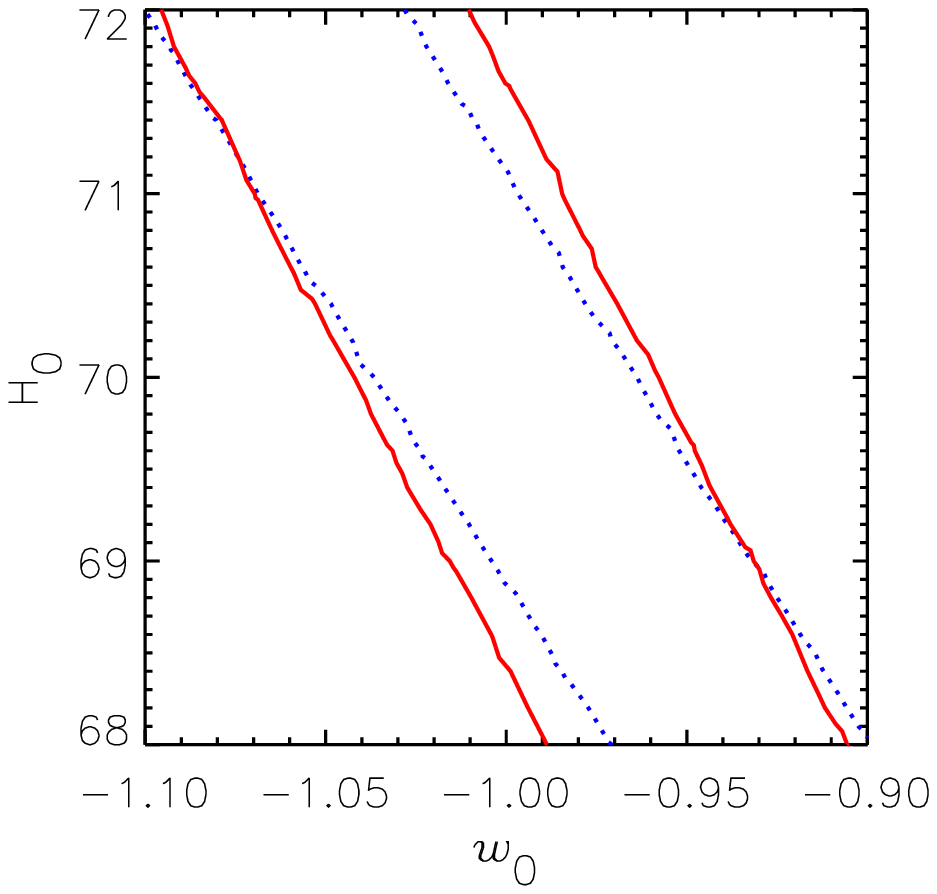}
\includegraphics[width=4.4cm]{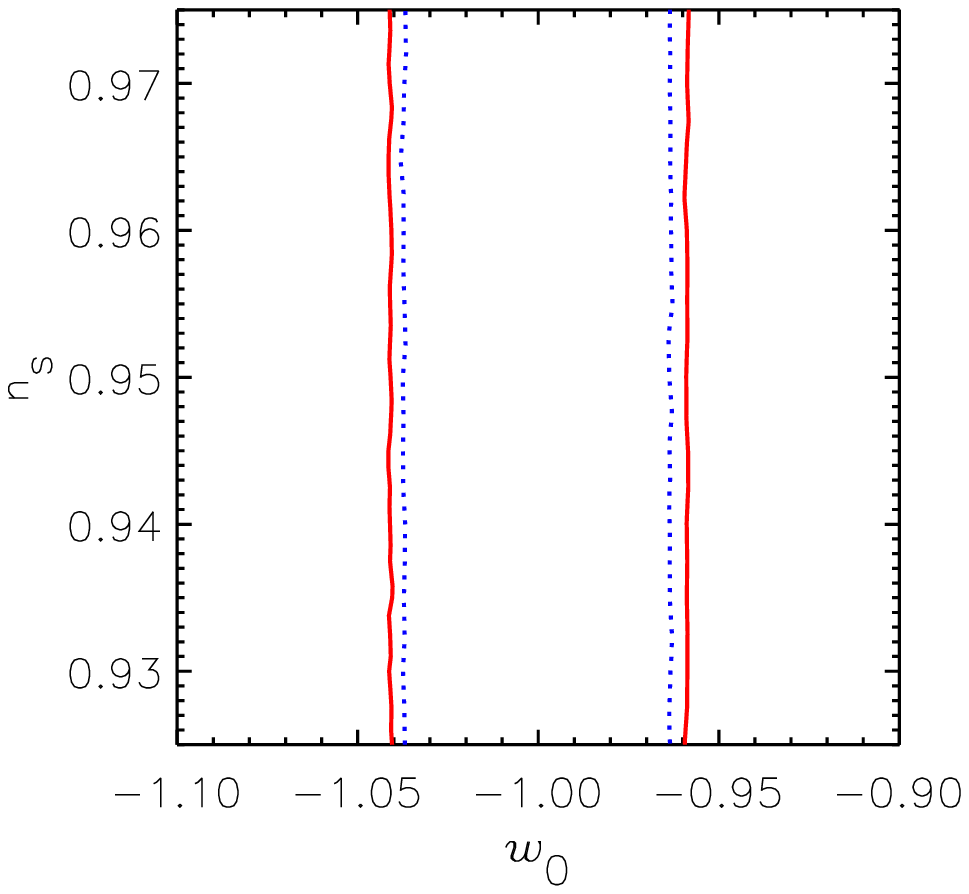}
\caption{\footnotesize Same as in Fig. \ref{fig:deg} but for $w_0$.}
\label{fig:deg_w0}
\end{figure}

More interesting are the dependences of the equation of state $w$ on the
cosmological parameters, in particular the dependence on $\omega_{\rm m}$ and $H_0$,
which enter the relation between the scaling factors and $w$. For a constant
$w$ model ($w=w_0$), we indicate the derived correlations in Fig. \ref{fig:deg_w0}, where
the solid line corresponds to $z=3$ and the dotted line to $z=1$. In contrast
to the previous case, these correlations are redshift dependent, for example
at $z=1$ the dependence of both the sound horizon and the Hubble parameter
$H(z)$ on $\omega_{\rm m}$ almost exactly cancel each other, whereas at $z=3$ the
effect originating in the Hubble parameter $H(z)$ is stronger.
Among the cosmological parameters the largest uncertainty in $w_0$ originates in
$H_0$, in particular at lower redshifts.

\subsection*{Light-cone effect}
We have already mentioned that the power spectra derived from the light cone output 
and corresponding snapshot at the mean redshift are almost 
on top of each other. The more evolved parts of the light-cone sample 
compensate almost precisely the less evolved parts, such that its power spectrum 
is almost identical to that at the mean redshift (see Fig. \ref{fig:ps_ratio}). 
In Fig. \ref{fig:lc_vs_sn}
we show the corresponding PDFs of the scaling factors. The lines show the $68\%$ and $95\%$ confidence levels obtained 
from the dark matter snapshot (dashed) and light-cone (dotted) power spectra 
at redshift $z=3$ (left) and $z=1$ (right) in real (top) and redshift space 
(bottom). 
The differences in the error ellipses derived from the light-cone and snapshot
data are overall small. Hence, light-cone effects in this survey volume and at these redshifts are unimportant to 
our fitting.

\subsection*{Reconstruction}
\begin{figure}[hbt]
\centering
\includegraphics[width=4.4cm]{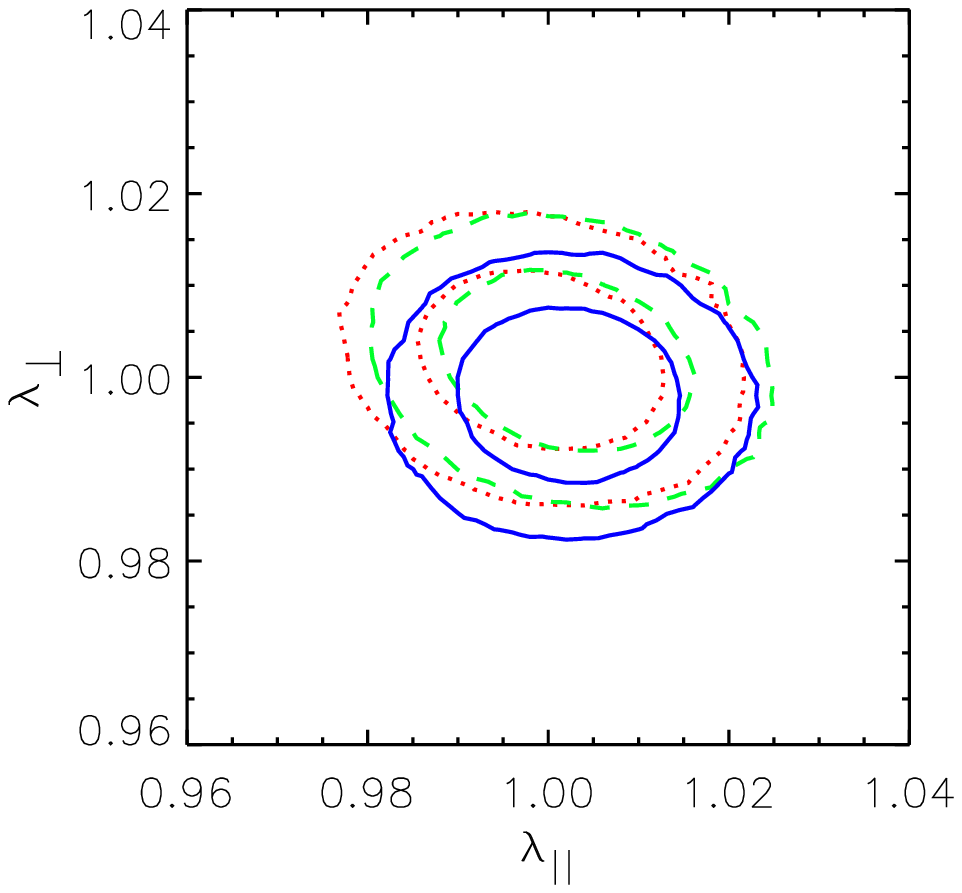}
\includegraphics[width=4.4cm]{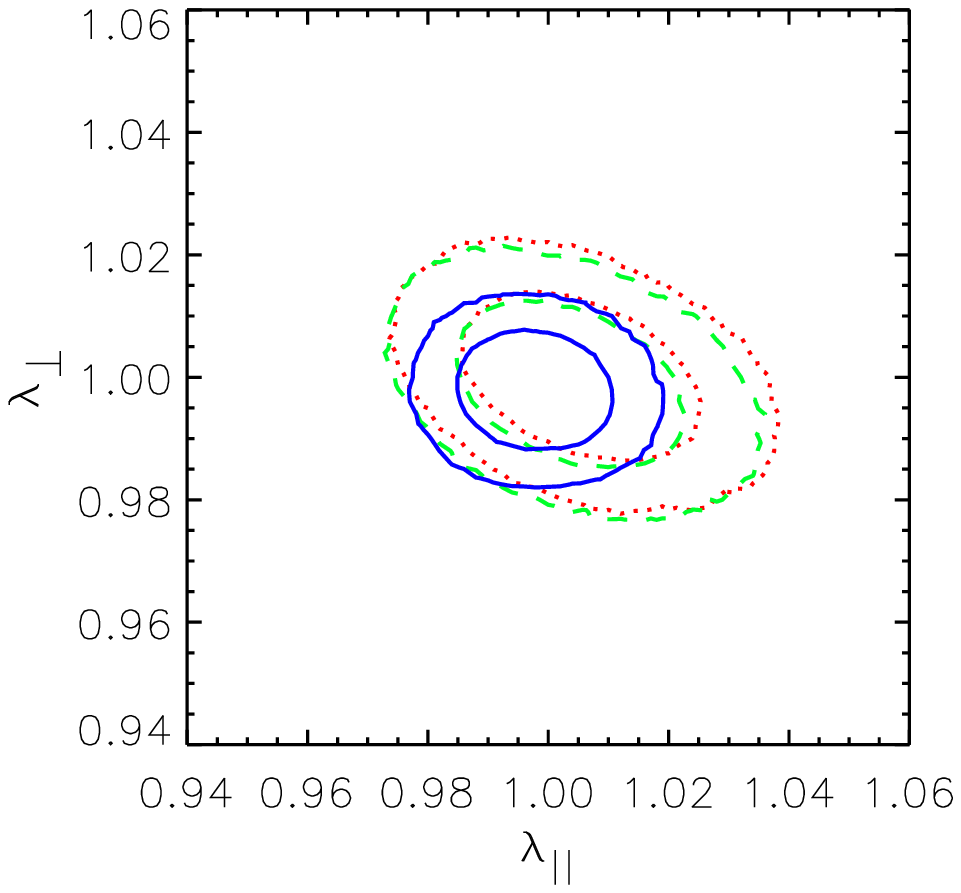}
\includegraphics[width=4.4cm]{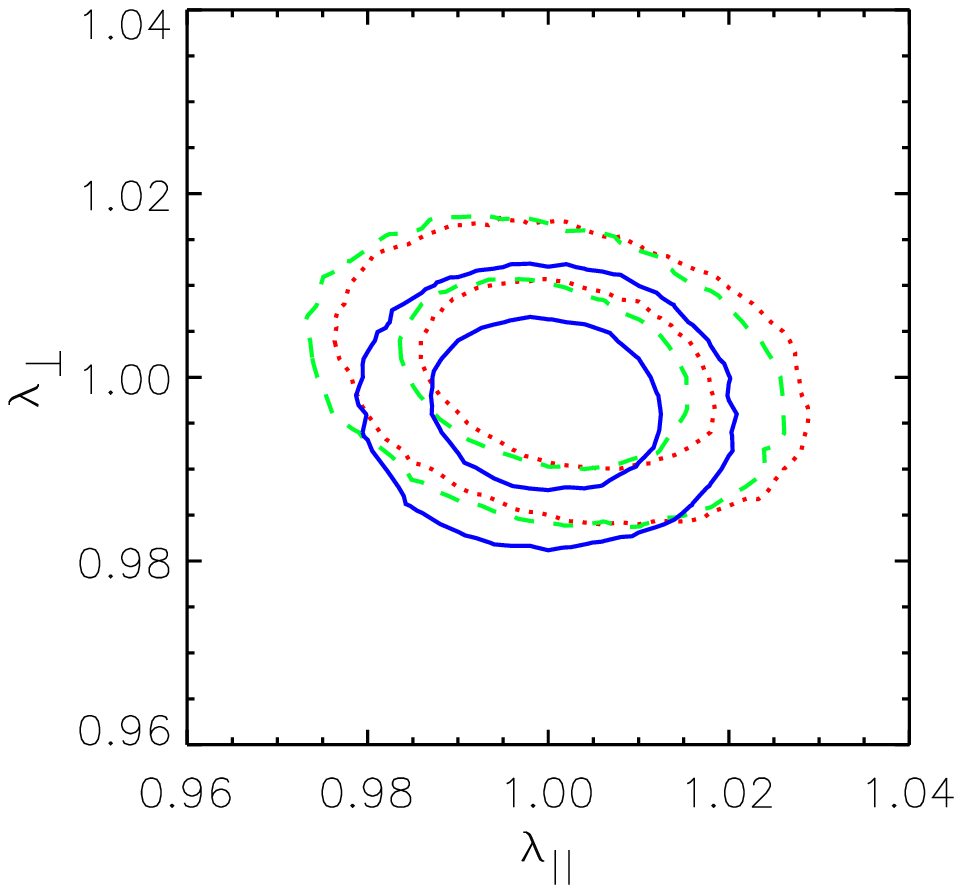}
\includegraphics[width=4.4cm]{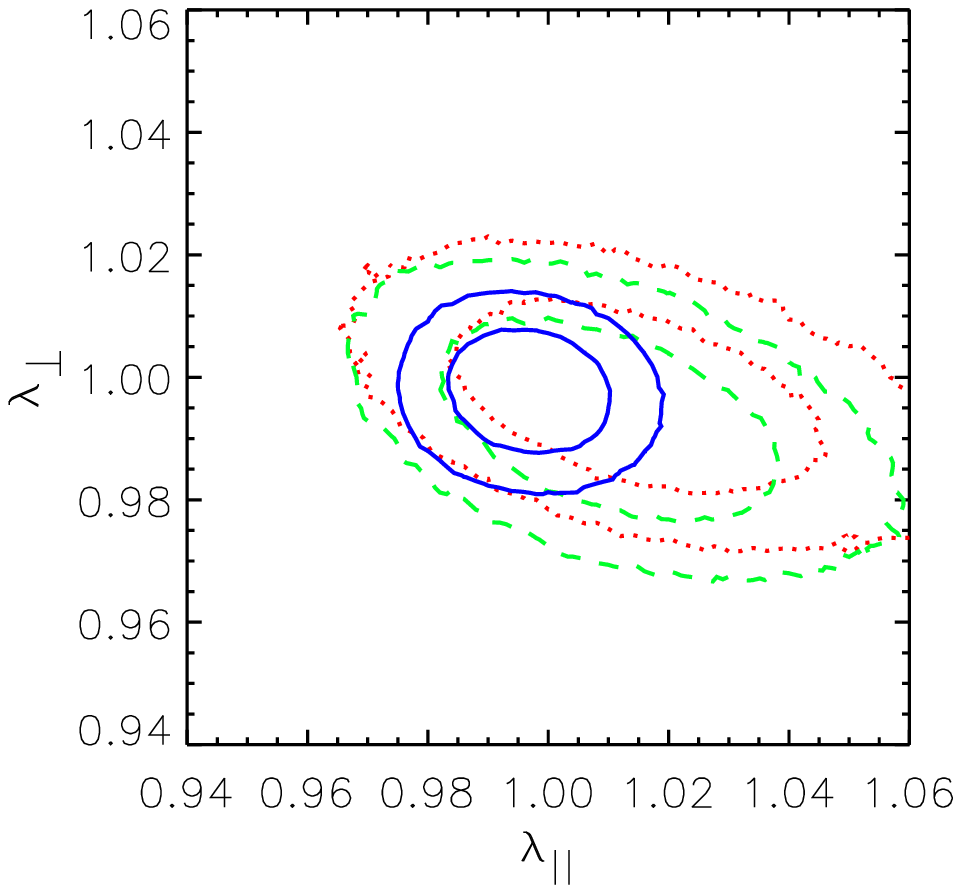}
\caption{\footnotesize The marginalized probability 
distribution functions (PDF) of the scaling factors at redshifts $z=3$ (left)
and $z=1$ (right) in real space (top) and redshift space (bottom) derived
from the light cone (dotted), snapshot (dashed), and 
``reconstructed'' light cone (solid).}
\label{fig:lc_vs_sn}
\end{figure}

The error ellipses for the reconstructed samples are shown in
Fig. \ref{fig:lc_vs_sn} as solid lines. We observe that for data at $z=1$ in
redshift space the reconstruction of the BAO shrinks the error ellipses by a
factor of three. For this reason, it would be desirable to develop a
reconstruction method that can be applied to noisy and biased density 
fields. For surveys at redshift $z=3$, this effect is less important.

\subsection*{ Orientation of the error ellipse}
The errors in the scaling factors originate in two different sources. One source is
the errors in the power spectrum; the other source is the
uncertainties in the cosmological parameters $\omega_{\rm m}$ and $\omega_{\rm
  b}$, which produce an uncertainty in the sound horizon, i.e.~the scale of
the BAO. 
If the principal error could be attributed to the uncertainty in the physical 
sound horizon, the orientation of the ellipse would be approximately in the
direction of the line defined by
$\lambda_\bot=\lambda_{||}$. For a low signal-to-noise
ratio power spectrum for which the scale of the BAO in the power spectrum 
cannot be determined accurately, the orientation of the error ellipse is
instead in the direction of the line of $\sqrt[3]{\lambda_\bot^2\lambda_{||}}=\rm const$. 

\subsection*{Cosmological distortion}
We assess if the approximations to compensate for an incorrect reference 
cosmology i.e.~the scaling relations given in Eq. (\ref{approx}), are sufficiently 
accurate for our simulated surveys. We calculate the distances 
assuming three reference cosmologies that differ only in terms of $w_{\rm ref}$ 
and compute the corresponding power spectra.

\begin{figure}[htbp]
\centering
\includegraphics[angle=-90,width=9cm]{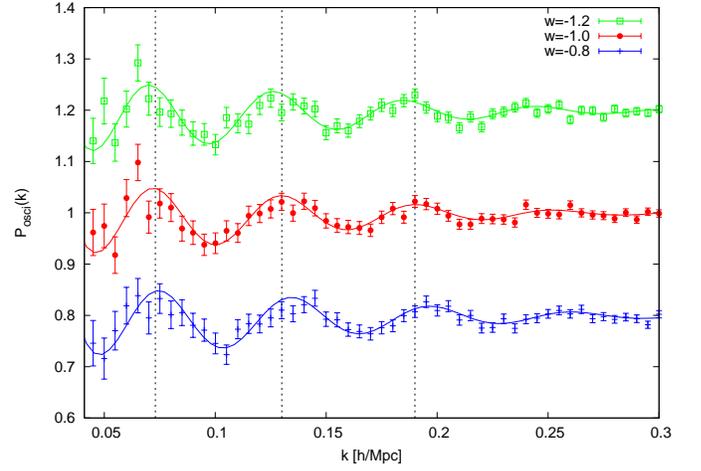}
\caption{\footnotesize The oscillatory part of 
the real space power spectra at redshift $z=3$ derived for three 
different cosmologies and its best-fit function. For display purposes, the 
power spectrum for $w_{\rm ref}=-1.2(-0.8)$ is shifted $+0.2(-0.2)$ along the 
$y$-axis. }
\label{fig:w_osci}
\end{figure}

As an example, the oscillatory parts of the dark matter power spectra at $z=3$ 
in real space are shown in Fig. \ref{fig:w_osci}. We observe 
that the scale of the oscillations is compressed and stretched for 
$w_{\rm ref}=-1.2$ and $w_{\rm ref}=-0.8$, respectively.

\begin{figure}[htbp]
\centering
\includegraphics[angle=-90,width=9cm]{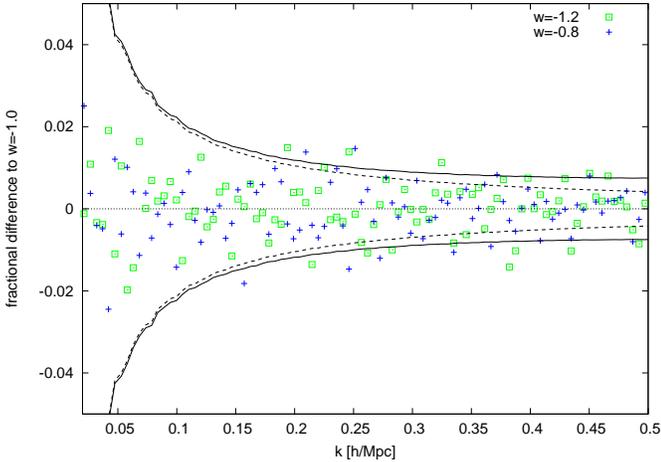}
\caption{\footnotesize The fractional difference of the scaled real-space 
power spectra at redshift $z=3$ is shown for the three different reference 
cosmologies. Note that the data points for $w=-1.2$ and $w=-0.8$ are shifted 
vertically by $+0.0025$ and $-0.0040$, respectively, to enhance the visibility 
of the scatter. The lines show the intrinsic error in the power
spectrum due to cosmic variance (dashed line) plus shot noise (solid line).}
\label{fig:w_full}
\end{figure}

We scale the one-dimensional power spectra according to the scaling 
relation, i.e.~$P(k)\rightarrow \lambda_{\rm  iso}^3\,P(k/\lambda_{\rm iso})$, 
where the factor $\lambda_{\rm iso}^3$ is due to the scaled volume element.
The fractional difference of these scaled power spectra  with respect 
to that of the correct cosmology, i.e.~$w=-1$, is shown in Fig. \ref{fig:w_full}. 
The data points for $w=-1.2$ (squares) and $w=-0.8$ (plus signs) are shifted 
by $+0.25\%$ and $-0.40\%$, respectively, to center them on the zero line. 
These small shifts have a similar origin as in the light-cone versus snapshot 
comparison; we calculate the scaling factor at the mean redshift, which is 
not equal to the mean scaling factor. The noise in the scaled power 
spectra is substantially smaller on large scales than the error, which is 
already implicit in the power spectrum due to cosmic variance (dashed line) 
and additional shot noise (solid line). To determine if this 
(additional) noise impairs the fitting of the BAO and
with it the measurement of the equation of state $w$, we fit the corresponding data of the 
twelve $256^3$ simulations by assuming a constant $w$. We find that the best-fit 
value for $w_0$ varies among the different reference cosmologies for a 
single simulation by not more than $4\%$, and is typically $\sim 2\%$. This
should be compared with the standard deviation for a single $w_0$ measurement, 
which is about $3\%$. All PDFs scatter about the mean value of 
$w_0=-1$, no systematic effects are noticeable. We conclude that the scaling 
relations do not introduce a noticeable bias or enlarge the error in $w$. 
For real data, a consistency check would be  
to recalculate the distances and corresponding power spectrum and 
redo the analysis using the measured $w$ as the reference 
value for the assumed cosmology.

\subsection*{Constraining $w$ with the (an)isotropic power spectrum}
\begin{figure}[htbp]
\centering
\includegraphics[width=4.4cm]{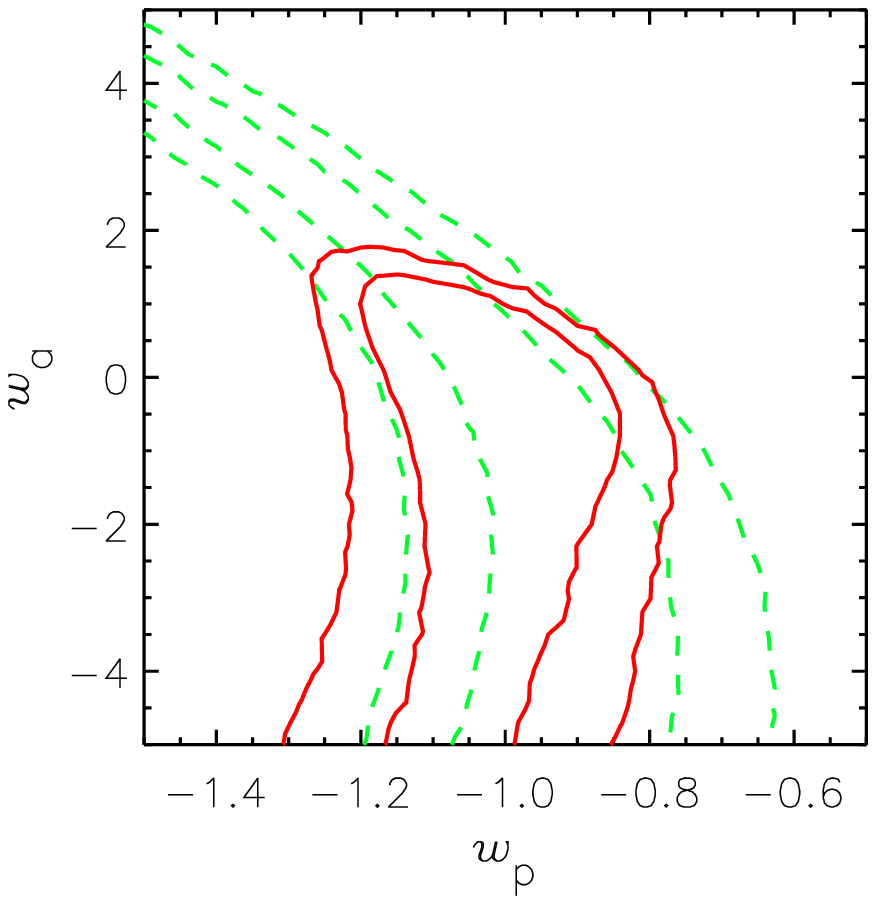}
\includegraphics[width=4.4cm]{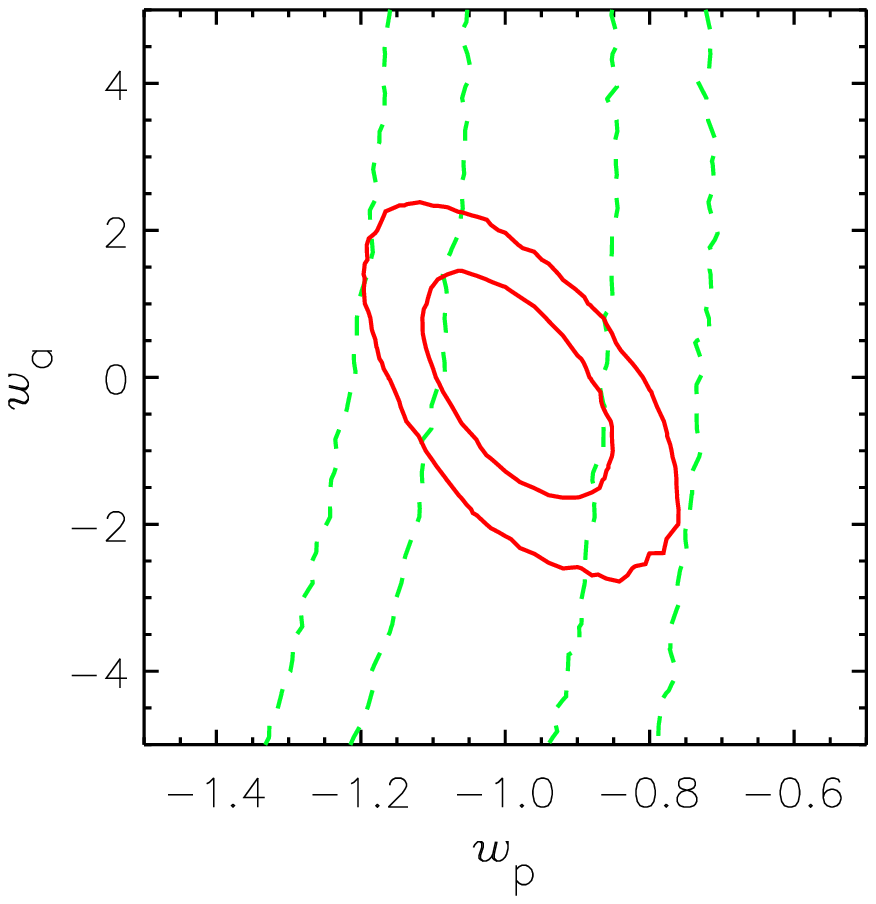}
\caption{\footnotesize This plot shows the $68\%$ and $95\%$ confidence levels 
for the redshift-dependent $w$ model obtained from the one-dimensional 
(dashed lines) and two-dimensional (solid lines) dark matter power spectrum 
at redshift $z=3$ (left) and $z=1$ (right) in real space.}
\label{fig:2dvs1d}
\end{figure}

The two-dimensional power spectrum provides the possibility to measure 
$\lambda_\parallel$ and $\lambda_\bot$, instead of the combination 
$\lambda_{\rm iso}$ only, as in the case of the one-dimensional power spectrum. 
Since both scaling factors depend on $w$ in different ways, we expect that $w$
is more accurately constrained by the two-dimensional 
power spectrum. In the case of a simple constant $w$, this is not the case:
both the mean values and the errors are similar to those derived from the
one-dimensional power spectra. For more complex models of $w$, the measurement of
$\lambda_\parallel$ and $\lambda_\bot$ is very helpful as we see from
Fig. \ref{fig:2dvs1d}, where we applied the model $w=w_0+(1-a) w_a$.  
For display purposes, we performed a coordinate transformation of the
parameters to the ``pivot'' system
$w_0+(1-a) w_a=w_{\rm p}+(a_{\rm p}-a) w_a$ where we chose $z_{\rm p}=0.3$ as discussed below. 
The left (right) panel indicates the constraints obtained from the dark
matter light-cone power spectrum around $z=3$ ($z=1$) in real space. One sees 
that for $z=3$ the contour lines are open towards negative $w_a$ even for the 
two-dimensional case. Nevertheless, in both redshift cases the use of the 
anisotropic power spectrum tightens the constraints substantially.

\subsection*{Estimates for future surveys}
\begin{figure}[htbp]
\centering
\includegraphics[width=4.4cm]{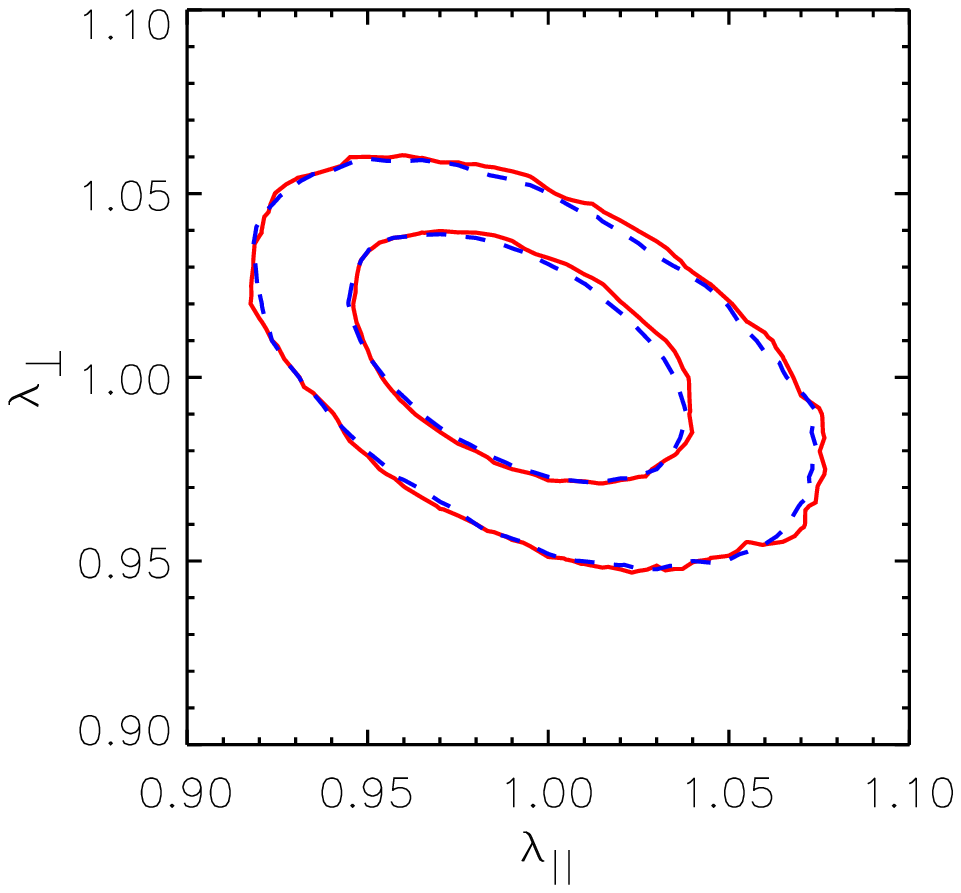}
\includegraphics[width=4.4cm]{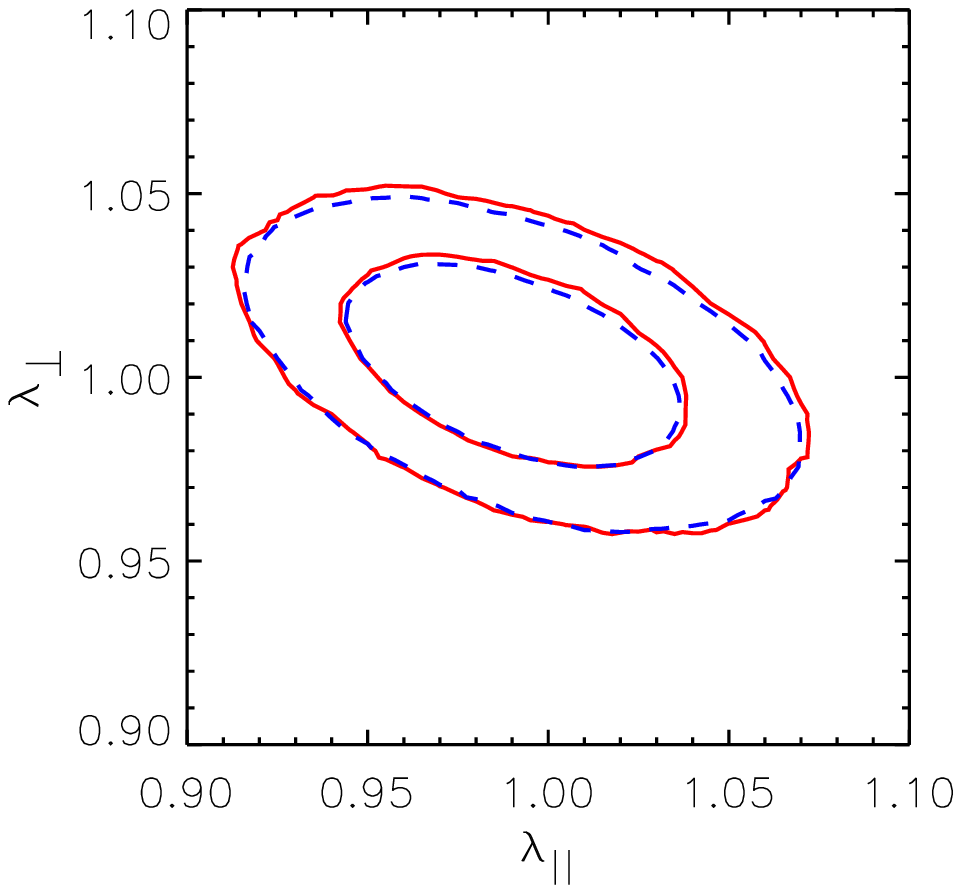}
\includegraphics[width=4.4cm]{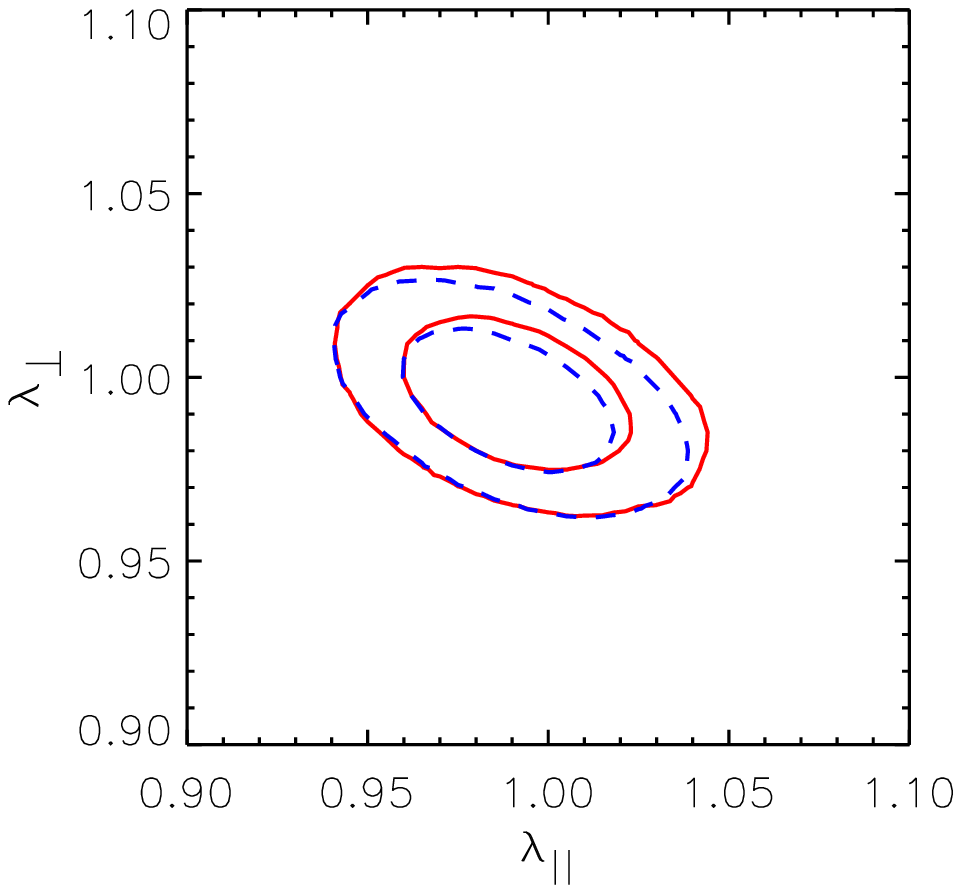}
\includegraphics[width=4.4cm]{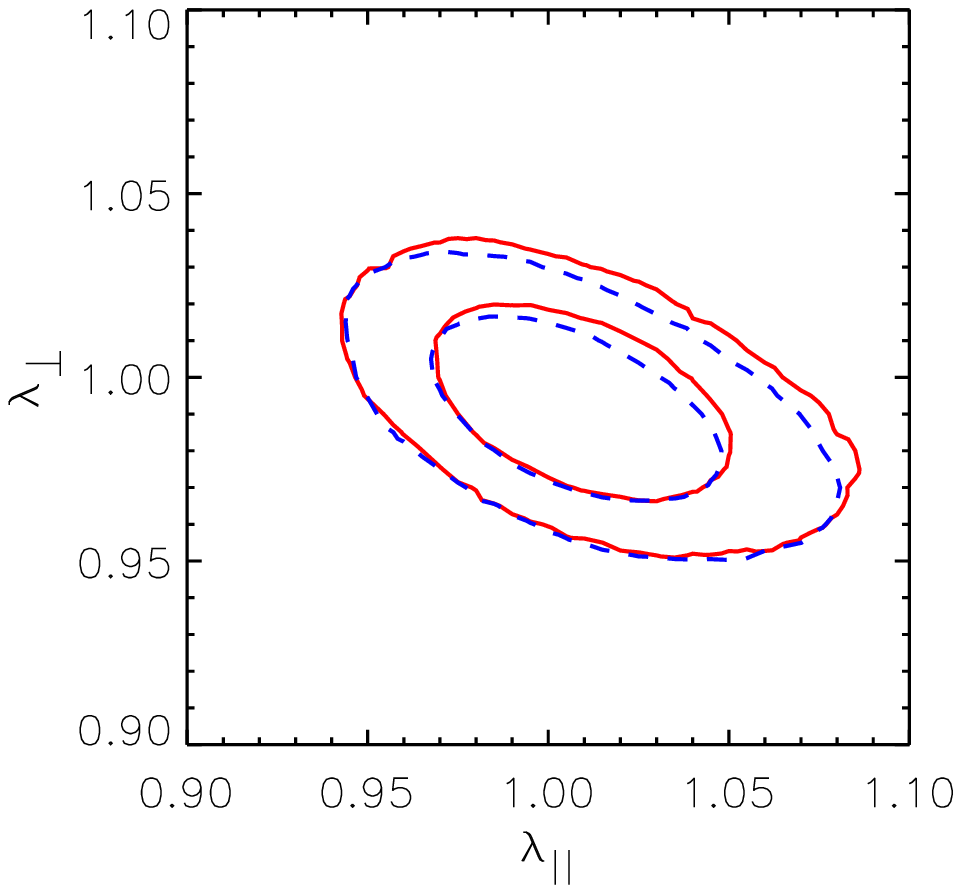}
\caption{\footnotesize Typical contour plots of the joint PDF of
$\lambda_\parallel$ and $\lambda_\bot$ obtained from the mock catalogs at
redshift $z=3$ (left: weak (top) and strong (bottom) bias) and $z=1$ 
(right: 1 million (top) and 2 million (bottom) galaxies) are shown. 
The contour lines correspond to 68\% and 95\% confidence level.  The dashed 
line shows the results when the cosmological parameters are fixed at the 
values of the simulation.}
\label{fig:2d_gal}
\end{figure}

For a prediction of how well upcoming observations will enable $w$ to be measured, 
we analyze the mock galaxy catalogs introduced above, namely the weakly and
strongly biased one million galaxy catalogs at redshift $z=3$ and the one and
two million galaxy catalogs at $z=1$. In Fig. \ref{fig:2d_gal}, we present
typical fitting results of the scaling factors for each type of
catalogs. For the solid lines, we used the aforementioned priors on the
cosmological parameters, whereas for the dashed lines we fixed these parameters.

\begin{figure}[htbp]
\centering
\includegraphics[angle=-90,width=4.3cm]{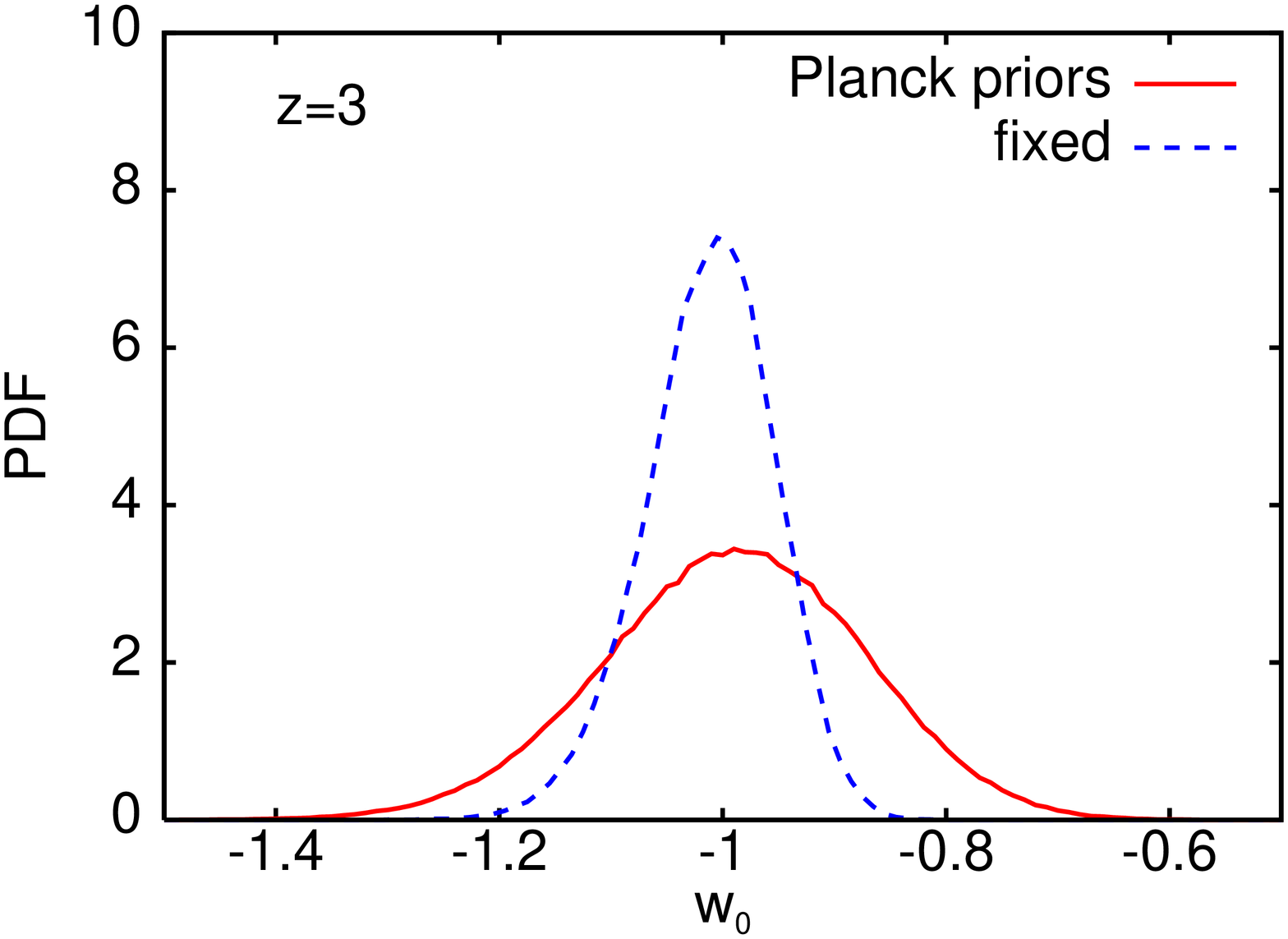}
\includegraphics[angle=-90,width=4.3cm]{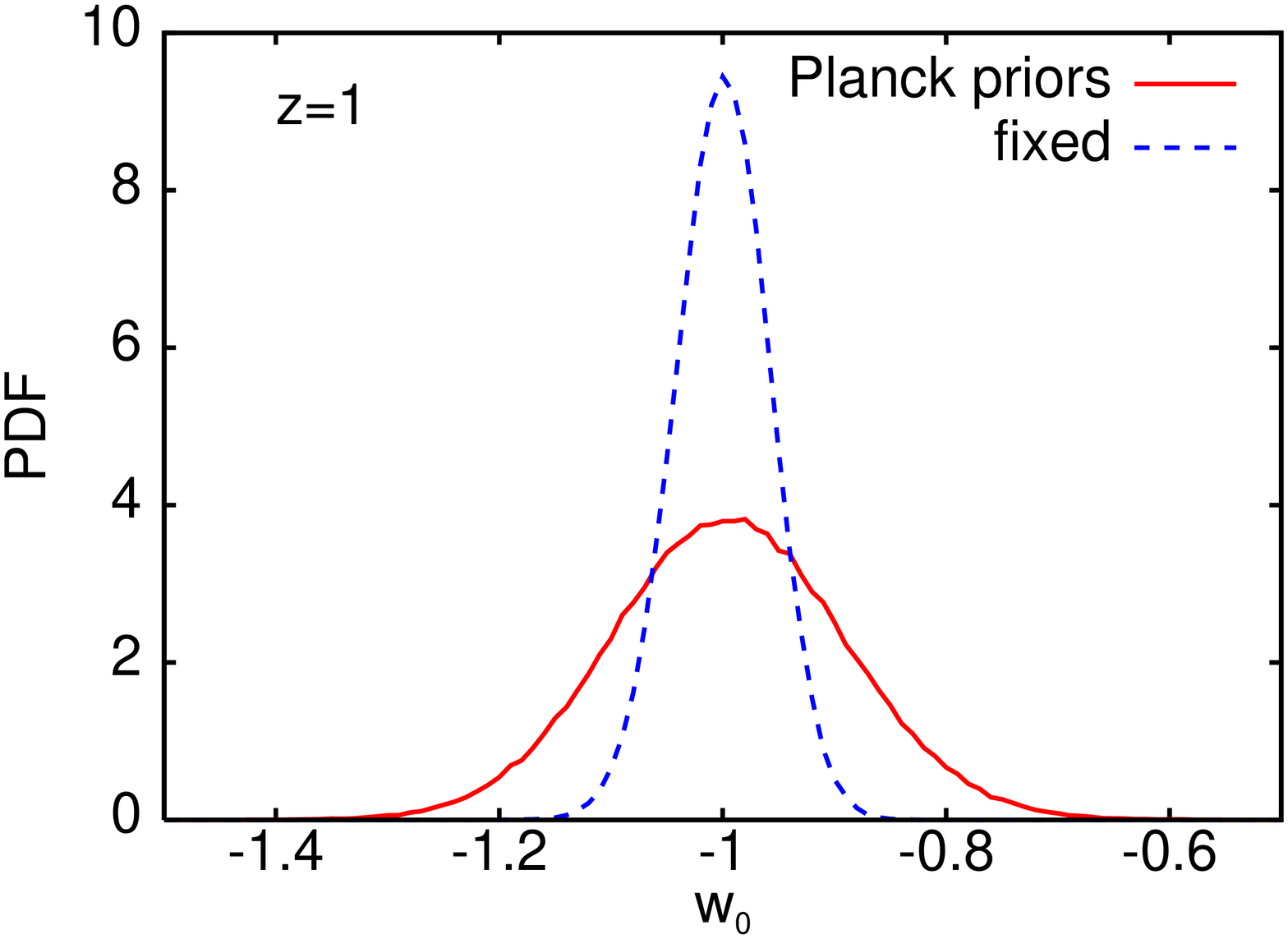}
\caption{\footnotesize PDFs for constant $w$ derived from the one million
  strongly biased galaxy sample at $z=3$ (left) and from the two million
  galaxy at $z=1$ (right).}
\label{fig:w0_gal}
\end{figure}

Since the error in the scaling factors is dominated by the errors in the power
spectrum, improving the accuracy of the cosmological parameters to higher than
that of the assumed priors does not significantly reduce the error in the scaling
factors. However, it tightens the constraints on $w$ (see Fig. \ref{fig:w0_gal} for
constant $w$): the uncertainty in $H_0$, in particular, degrades the measurement of $w$.

For the strongly biased sample at $z=3$ of one million galaxies in 
a volume of $10\, {\rm Gpc}^3$, the mean\footnote{We used 10 samples to derive 
the mean values.} uncertainties in the Hubble 
parameter ($H(z)\varpropto \lambda_\parallel)$ and angular diameter 
distance ($D_A(z) \varpropto 1/\lambda_\bot$) are $1.8\%$ and 
$1.2\%$ ($68\%$ c.l.), respectively. This corresponds to an error of $\sim 12\%$ 
for a constant $w$. By keeping the cosmological parameters fixed, the 
uncertainty in $w$ is lowered to $\sim 5\%$.

The corresponding numbers for the two million galaxy mock catalog at $z=1$ 
are $2.8\%$ and $1.6\%$ for the Hubble parameter and angular diameter
distance, respectively. In this case, we derive an accuracy of $\sim 11\%$ and $\sim 4\%$ 
for a constant $w$ with Planck priors and fixed cosmology, respectively.

To translate the results for the scaling factors into constraints
of the redshift-dependent $w$-model $w=w_0+(1-a)w_a$, we combine the data 
of the strongly biased one million galaxy catalog at $z=3$ and two
million galaxy catalog at $z=1$ with 192 SN Ia observations 
\citep{davis,wv07,riess2007,astier}. In addition we include constraints from
CMB measurements by holding the ratio of the distance to the last
scattering surface and the sound horizon fixed: $D_A(z=z_{\rm lss})/s=\rm const$.

\begin{figure}[htbp]
\centering
\includegraphics[width=4.4cm]{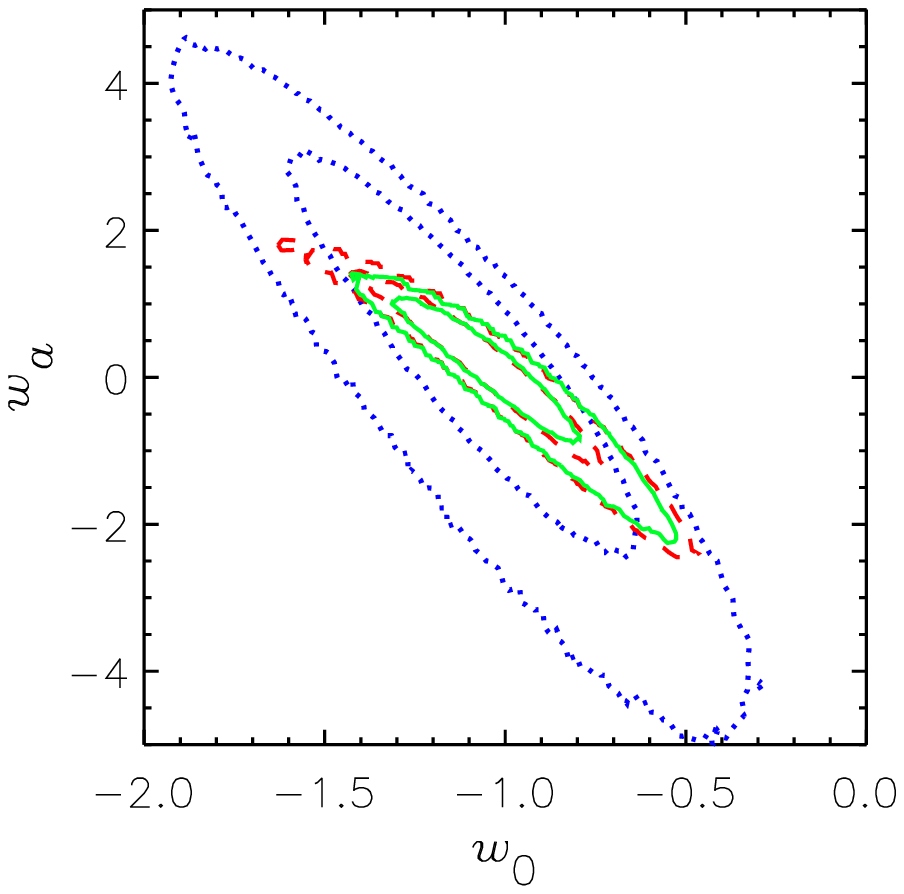}
\includegraphics[width=4.4cm]{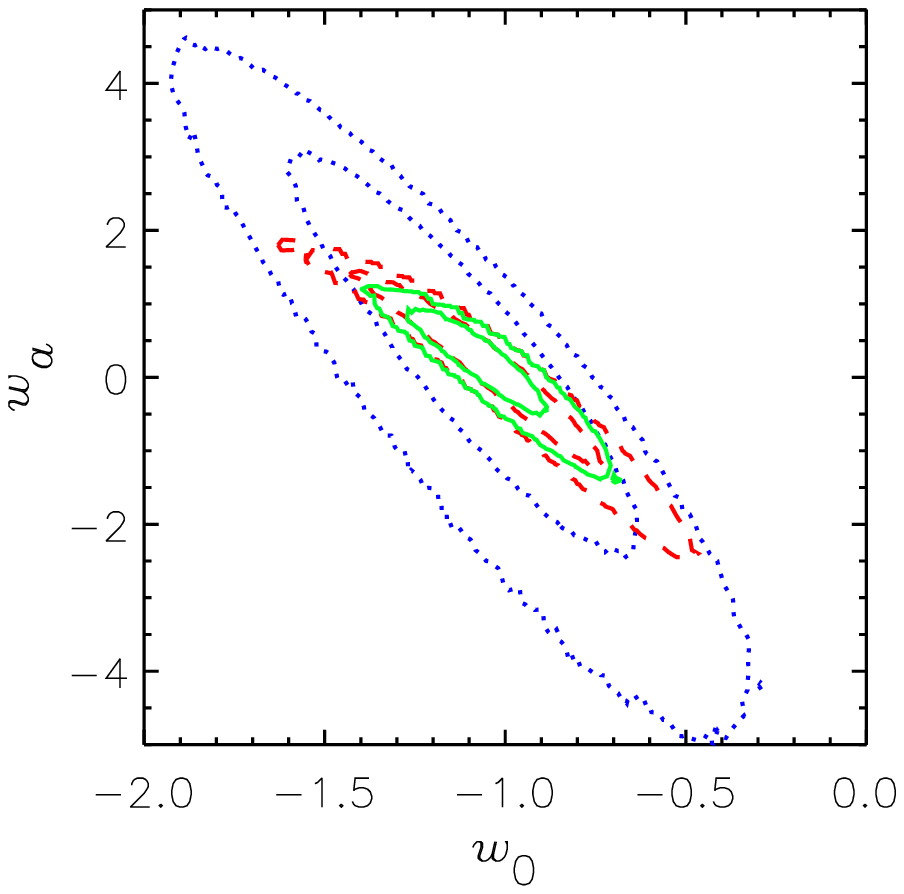}
\includegraphics[width=4.4cm]{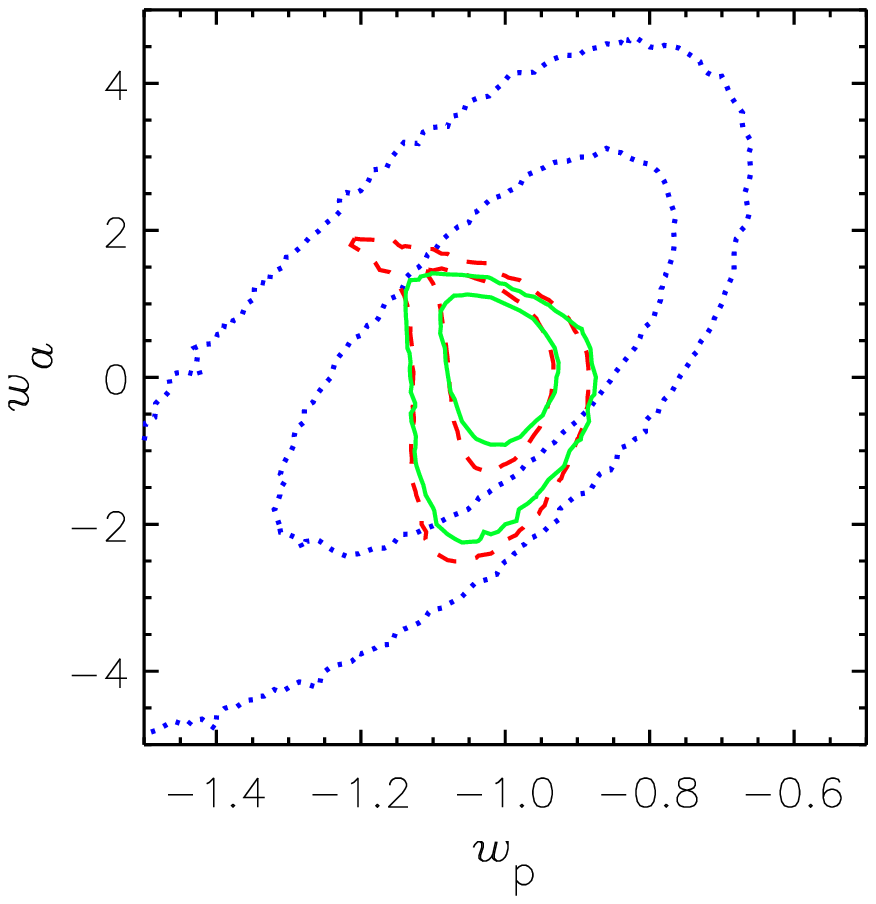}
\includegraphics[width=4.4cm]{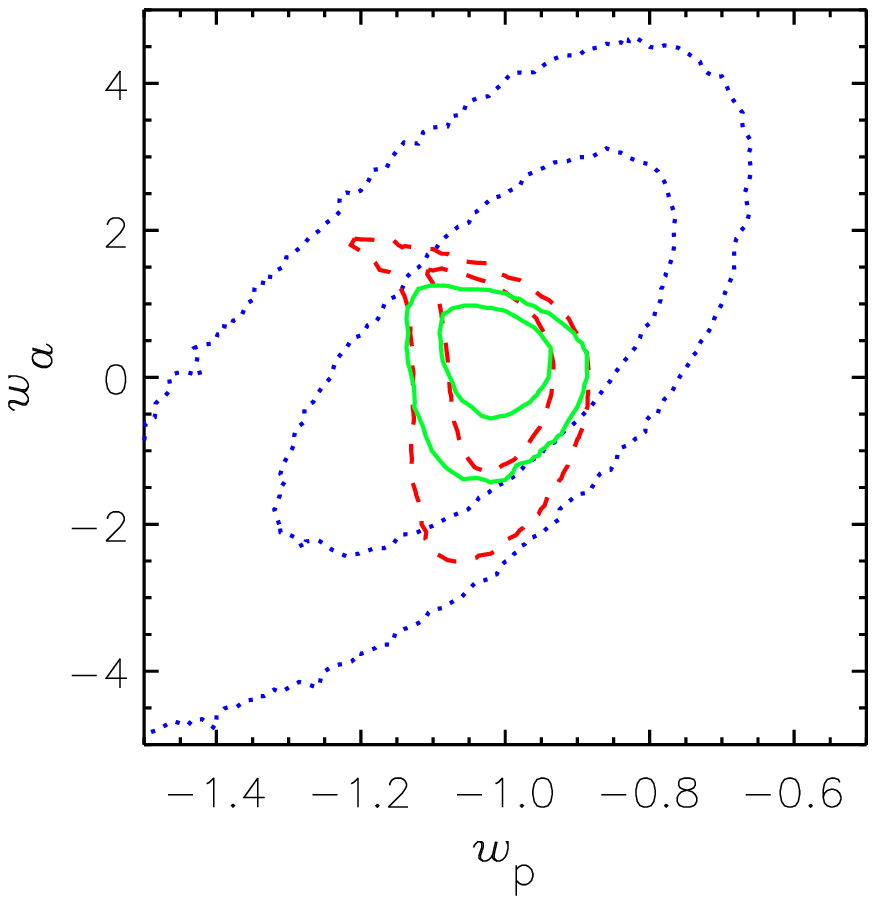}
\caption{\footnotesize The $68\%$ and $95\%$ confidence level for the
  parameters of the model $w=w_0+(1-a) w_a$ obtained from SN (dotted), SN+CMB
  (dashed), and SN+CMB+BAO (solid). Results for the strongly biased one
  million galaxy sample  are shown at $z=3$ (left) and for the weakly biased 
  two million galaxy sample at $z=1$ (right). 
  At the bottom, we show the same results  
  but transformed to the pivot system $z_{\rm p}=0.3$.}
\label{fig:w0wa}
\end{figure}

\begin{figure}[htbp]
\centering
\includegraphics[width=4.4cm]{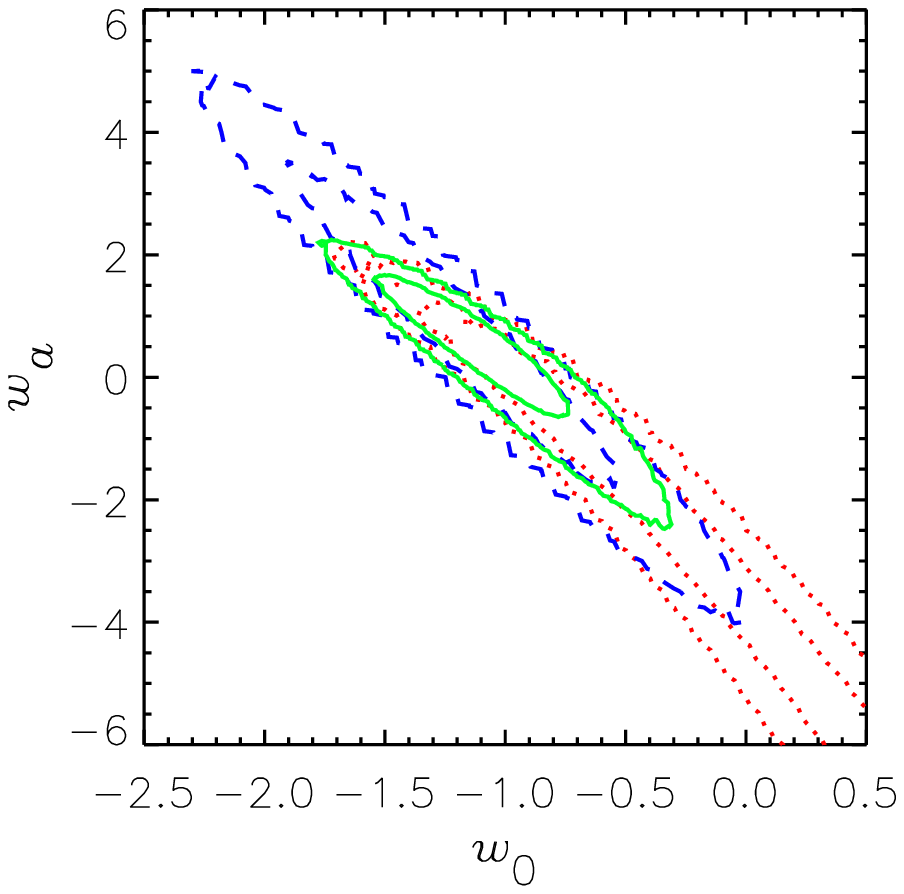}
\includegraphics[width=4.4cm]{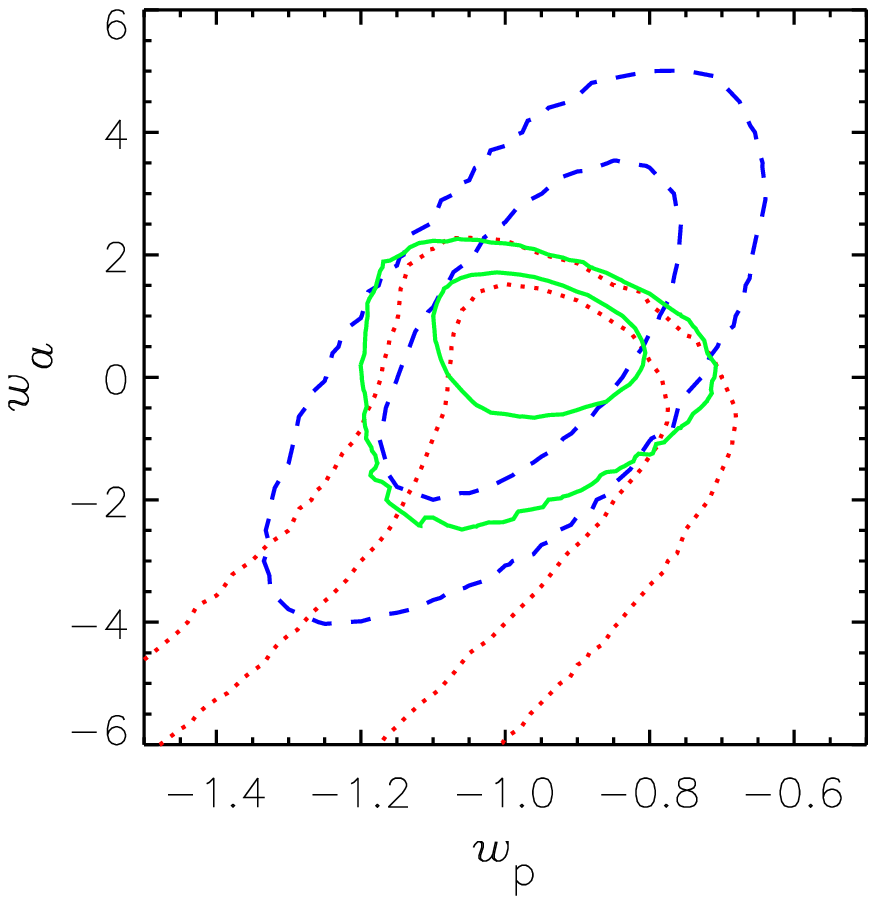}
\caption{\footnotesize The $68\%$ and $95\%$ confidence level for the
  parameters of the model $w=w_0+(1-a) w_a$ obtained from the BAO measurements in the strongly biased one
  million galaxy sample at $z=3$ (dotted) and in the weakly biased 
  two million galaxy sample at $z=1$ (dashed). The solid lines show the combination of both BAO measurements. On the right, the contours are transformed to the pivot system of the combined dataset $a_{\rm p}=0.7$.}
\label{fig:w0wa_bao}
\end{figure}

The confidence contours for $w_0$ and $w_a$ are shown in the upper panels of Fig. \ref{fig:w0wa}. In the lower panels, the confidence 
contours are presented in the pivot variable $w_{\rm p}=w_0+(1-a_{\rm  p})w_a$, 
where we choose $a_{\rm p}$ such that the error ``ellipse'' from the combined datasets (solid line) is least tilted; we find $a_{\rm p}=1/(1+0.3)$. 
One sees that the intersection angle 
of the BAO contours with respect to the error ellipse derived from SN data 
varies with redshift (see also Fig. \ref{fig:w0wa_bao}). 
For $z=3$, the orientation is very similar to the CMB contours, 
whereas for $z=1$, it is more aligned with the SN ellipse. Measurements at both 
redshifts would therefore constrain the parameters of this model significantly 
more than two (or twice a good) measurements at a single redshift.

In Fig. \ref{fig:w0wa_bao}, we show the constraints on $w$ derived alone from BAO measurements at two different redshifts, namely $z=1$ (dashed lines) and $z=3$ (dotted lines), and the combination of both measurements (solid lines). The redshift-dependent shape and orientation of the contour lines is clearly evident in the pivot system (right panel).

For all confidence contours in Fig. \ref{fig:w0wa} and Fig. \ref{fig:w0wa_bao}, the above stated priors for the cosmological parameters $\omega_{\rm m}$, $\omega_{\rm b}$, $n_{\rm s}$, and $H_0$ were used.

\section{Discussion and Conclusions}
We have simulated large redshift surveys by ``observing'' mock galaxies on a light
cone obtained from an N-body simulation with $512^3$ dark matter
particles in a $(1.5\, {{\rm Gpc}/h})^3$ box. By fitting the apparent 
scale of the BAO in the power spectrum, we have derived constraints on the 
EOS parameter $w$.

Our fitting method uses only the oscillatory part of the power spectrum and is
therefore insensitive to changes in the power spectrum due to nonlinear
evolution, redshift distortions, and scale-dependent galaxy bias. Fitting
methods that use the overall shape of the power spectrum suffer in general 
from these complicated physical effects (unless they are accurately 
modeled) and tend to provide biased results \citep{smith}.

The drawback of fitting only the oscillatory part is that it is insensitive to 
information provided by the overall shape of the power
spectrum. Determining the growth function would be
interesting in particular for dark energy constraints \citep{amendola1,amendola2}.

It is unclear how to extract the BAO from the power spectrum in the most
appropriate and accurate way. 
Our simple smoothing method works fairly well. It is essentially
parameter-free and produces unbiased and robust results.

We analyzed the degeneracies in the EOS $w$ with other cosmological
parameters. Given the expected accuracies in the cosmological parameters involved,
we found that more accurate measurements of $H_0$ would have the largest effect on lowering the
uncertainties in $w$, especially for observations at lower redshifts. 

Comparing the results of the light-cone power spectrum with 
those of the power spectrum of the snapshot at the corresponding 
redshift, we did not find evidence for substantial differences. For the surveys under consideration, light cone effects therefore do not play a role in determining $w$ with our fitting method.
This is in addition true, when we include redshift-dependent galaxy
bias. Fitting methods, sensitive to the overall shape of the power spectrum,
in particular to the growth function, must include explicitly the light-cone effects to produce unbiased results.

The PIZA technique \citep{piza}, which is used to reconstruct the linear 
density field and
thereby the amplitude of the BAO, is effective if the density field is
known to sufficient accuracy. In the case of substantial shot noise and
unknown galaxy bias, more sophisticated reconstruction techniques are required.

To investigate the cosmological distortion due to an incorrect reference
cosmology, we adopted three different reference cosmologies with different values for 
$w_{\rm ref}=-0.8$, $-1.0$,  and $-1.2$. Employing the usual scaling relations 
(see Eq. (\ref{scalingfactors})), we found in all three cases similar fitting
results within the margins of error. We conclude that even for this large survey 
volume the simple scaling of the power spectrum is fairly accurate. 
Nevertheless, for real data we propose to use an iterative scheme; 
after measuring $w$ for an assumed reference cosmology, one would then repeat 
the analysis for the updated reference cosmology, using the measured value of 
$w$ for $w_{\rm ref}$.

By fitting the two-dimensional power spectrum, we can determine independently both scaling factors. For a constant $w$, the 
two-dimensional fitting does not improve the constraints on $w$ significantly.
For models of $w$ that are not so tightly constrained, independent measurements of the Hubble 
parameter ($H(z)\varpropto \lambda_\parallel)$ and the angular diameter 
distance ($D_A \varpropto 1/\lambda_\bot$) will however be very helpful. For the model
$w=w_0+(1-a)w_a$, this is especially true at lower redshifts.

Our estimates for future surveys show that BAO measurements around redshifts
$z=3$ and $z=1$ in combination with SN and CMB data tighten the constraints on 
dynamical dark energy substantially and that these redshift surveys deliver 
complementary results. For a constant $w$-model the BAO measurements from our 
two mock catalogs at $z=3$ and $z=1$ alone will constrain $w_0$ to an accuracy of $\sim 12\%$ and $\sim 11\%$, respectively.

To achieve more realistic predictions, the survey geometry, i.e.~the 
window function, has to be taken into account. A more realistic galaxy 
bias scheme, tailored in particular for the target galaxies, should also be 
developed. For such a study, large-volume simulations of far higher mass resolution are needed.

\section*{Acknowledgments}
We are grateful to Jochen Weller for advice concerning the Monte Carlo Markov
chain sampling method and dark energy models. We thank Karl Gebhardt and
Eiichiro Komatsu for helpful discussion and useful comments on the draft.
We acknowledge useful and constructive remarks by the anonymous referee.

\bibliographystyle{aa}

\bibliography{paper}

\end{document}